\documentclass{pnastwo}
\usepackage{graphicx}

\usepackage{PNAStwoF}

\usepackage{amssymb,amsfonts,amsmath}

\contributor{Submitted to Proceedings
of the National Academy of Sciences of the United States of America}
\url{www.pnas.org/cgi/doi/10.1073/pnas.0709640104}
\copyrightyear{2008}
\issuedate{Issue Date}
\volume{Volume}
\issuenumber{Issue Number}

\begin{document}

\title{Size distribution of particles in Saturn's rings from aggregation and fragmentation}

\author{Nikolai Brilliantov\affil{1}{University of Leicester, Leicester, United Kingdom},
P. L. Krapivsky\affil{2}{Department of Physics, Boston University, Boston, USA},
Anna Bodrova\affil{3}{Moscow State University, Moscow, Russia}\affil{4}{University of Potsdam, Potsdam, Germany},
Frank Spahn\affil{4}{},
Hisao Hayakawa\affil{5}{Yukawa Institute for Theoretical Physics, Kyoto University, Kyoto, Japan},
Vladimir Stadnichuk\affil{3}{},
\and J\"urgen Schmidt\affil{6}{Astronomy and Space Physics, University of Oulu, Oulu, Finland}\affil{4}{}}

\contributor{Submitted to Proceedings of the National Academy of Sciences
of the United States of America}

\maketitle
\begin{article}

\begin{abstract}
Saturn's rings  consist of a huge number of water ice particles, with a tiny addition of rocky material. They form a flat disk, as the result of an interplay of angular momentum conservation and the steady loss of energy in dissipative inter-particle collisions. For particles in the size range from a few centimeters to a few meters, a power-law distribution of radii, $\sim r^{-q}$ with $q \approx 3$, has been inferred; for larger sizes, the distribution has a steep cutoff. It has been suggested that this size distribution may arise from a balance between aggregation and fragmentation of ring particles, yet neither the power-law dependence nor the upper size cutoff have been established on theoretical grounds. Here we propose a model for the particle size distribution that quantitatively explains the observations. In accordance with data, our model predicts the exponent $q$ to be constrained to the interval $2.75 \le q \le 3.5$. Also an exponential cutoff for larger particle sizes establishes naturally with the cutoff-radius being set by the relative
frequency of aggregating and disruptive collisions. This cutoff  is much smaller than the typical scale of micro-structures seen in Saturn's rings.
\end{abstract}


\dropcap{B}ombardment of Saturn's rings by interplanetary meteoroids \cite{cuzzi1990,cuzzi1998,tiscareno2013} and the observation of rapid processes in the ring system \cite{cuzzi2010} indicate that the shape of the particle size distribution is likely not primordial or a direct result from the ring creating event. Rather, ring particles are believed to be involved in an active accretion-destruction dynamics~\cite{harris1975,davis1984,weidenschilling1984,GorkavyiFridman1985,longaretti1989,canup1995,SpahnetalEPL:2004,esposito2006,icarus2012} and their sizes vary over a few orders of magnitude as a power-law~\cite{marouf1983,zebker1985,cuzzi2009,french2000b}, with a sharp cutoff for large sizes~\cite{zebker1983,tiscareno2006,Sremcevic2007,tiscareno2008}. Moreover, tidal forces fail to explain the abrupt decay of the size distribution for house-sized particles \cite{Guimaraes}.  One would like to understand: (i) can the interplay between aggregation and fragmentation lead to the observed size distribution, and (ii) is this distribution peculiar for Saturn's rings, or is it universal for planetary rings in general? To answer these questions quantitatively, one needs to elaborate a detailed model of the kinetic processes in which the rings particles are involved. Here we develop a theory that quantitatively explains the observed properties of the particle size distribution and show that these properties are generic for a steady-state, when a balance between aggregation and fragmentation holds. Our model is based on  the hypothesis that collisions are binary and that they may be classified as aggregative, restitutive or disruptive (including collisions with erosion); which type of collision is realized depends on the relative speed of colliding particles and their masses. We apply the kinetic theory of granular gases \cite{borderies1985,brillOUP} for the multi-component system of ring particles to quantify the collision rate and the type of collision.

\section{Significance}
{\bf Although it is well accepted that the particle size distribution in Saturn's rings is not primordial, it remains unclear whether the observed distribution is unique or universal. That is, whether it is determined by the history of the rings and details of the particle interaction, or if the distribution is generic for all planetary rings. We show that a power-law size distribution with large-size cutoff, as observed in Saturn's rings, is universal for systems where a balance between aggregation and disruptive collisions is steadily sustained. Hence, the same size distribution is expected for any ring system where collisions play a role, like the Uranian rings, the recently discovered rings of Chariklo and Chiron, and possibly rings around extrasolar objects.}

\section{Results and Discussion}
\subsection{Model}
 Ring particles may be treated as aggregates\footnote{The concept of aggregates  as Dynamic Ephemeral Bodies (DEB) in rings has been proposed in~\cite{weidenschilling1984}.} built up of primary grains~\cite{longaretti1989} of a certain  size $r_1$ and mass $m_1$.\footnote{Observations indicate that particles below a certain radius are absent in dense rings \cite{cuzzi2009}.} Denote by $m_k=km_1$ the mass of ring particles of "size" $k$ containing $k$ primary grains, and by $n_k$ their density. For the purpose of a kinetic description we assume that all particles are spheres; then the radius\footnote{In principle,  aggregates can be fractal objects, so that  $r_k \sim k^{1/D}$, where $D\leq 3$ is the fractal dimension of aggregates. For dense planetary rings it is reasonable to assume that aggregates are compact, so $D=3$.} of an agglomerate containing $k$ monomers is $r_k = r_1 k^{1/3}$. Systems composed of spherical particles may  be described in the framework of the Enskog-Boltzmann theory~\cite{araki1986,araki1988,araki1991}. In this case the rate of binary collisions depends on particle dimension and relative velocity. The cross-section for the collision of particles
of size $i$ and $j$ can be written as $\sigma^2_{ij}=\left(r_i+r_j\right)^2=r_1^2 (i^{1/3} + j^{1/3})^2$. The relative speed (on the order of $0.01 - 0.1 \, {\rm cm/s}$ \cite{cuzzi2009}) is determined by the velocity dispersions $\left< {\bf v}_i^2 \right>$ and  $\left< {\bf v}_j^2 \right>$ for particles of size $i$ and $j$. The velocity dispersion quantifies the root mean square deviation of particle velocities from the orbital speed ($\sim 20 \, {\rm km/s}$). These deviations follow a certain distribution, implying a range of inter-particle impact speeds, and thus, different collisional outcomes. The detailed analysis of an impact shows that for collisions at small relative velocities, when the relative kinetic
energy is  smaller than a certain threshold energy, $E_{\rm agg}$, particles stick together to form a joint
aggregate~\cite{DominikTielens1997,SpahnetalEPL:2004,Wada2009}. This occurs because adhesive forces acting
between ice particles' surfaces are strong enough to keep them together. For larger velocities, particles
rebound with a partial loss of their kinetic energy. For still larger impact speeds, the relative kinetic
energy exceeds the threshold energy for fragmentation, $E_{\rm frag}$, and particles break into
pieces~\cite{Wada2009}.

Using kinetic theory of granular gases one can find the collision frequency for all kinds of collisions and the respective rate coefficients: $C_{ij}$ for collisions leading to merging and $A_{ij}$ for disruptive collisions. The coefficients $C_{ij}$ give the number of aggregates of size $(i+j)$ forming per unit time in a unit volume as a result of aggregative collisions involving particles of size $i$ and  $j$. Similarly, $A_{ij}$ quantify disruptive collisions, when particles of size $i$ and $j$ collide and break into smaller pieces. These rate coefficients depend on masses of particles, velocity dispersions and threshold energies, $E_{\rm agg}$ and $E_{\rm frag}$:
\begin{eqnarray}
  \label{eq:KinCoef}
&&C_{ij}= \nu_{ij} \left(1-\left( 1+ B_{ij} E_{\rm agg} \right) \exp \left(- B_{ij} E_{\rm agg} \right)\right)
\nonumber \\
&&A_{ij}=\nu_{ij}\exp \left( - B_{ij} E_{\rm frag} \right)  \nonumber \\
&&\nu_{ij}= 4\sigma_{ij}^2 \sqrt{ \pi \left( \left< {\bf v}_i^2 \right> + \left< {\bf v}_j^2 \right> \right)/6} \\
&& B_{ij}= 3\, \frac{m_i^{-1}+m_j^{-1}}{\langle v_i^2  \rangle + \langle v_j^2 \rangle} \nonumber \,.
\end{eqnarray}
These results follow from the Boltzmann equation which describes evolution of a system in terms of the joint size-velocity distribution function (see the section below and SI Appendix). The governing rate equations for the concentrations of particles of size $k$ read:
\begin{eqnarray}
\label{kinpow}
\frac{dn_k}{dt} &=& \frac{1}{2}\!\sum_{i+j=k}C_{ij}n_in_j - \sum_{i=1}^{\infty}C_{ki}n_in_k  \\
&-& \sum_{i=1}^{\infty}A_{ki}n_in_k \left(1-\delta_{k1}\right) + \sum_{i=1}^{k}n_i
\sum_{j=k+1}^{\infty}A_{ij}n_jx_k\left(j\right) \nonumber \\
&+& \frac{1}{2} \sum_{i,j\geq k+1}  A_{ij}n_in_j\left[ x_k(i)+x_k(j)\right] \, . \nonumber
\end{eqnarray}
The first term on the right-hand side of Eq.~(\ref{kinpow}) describes the rate at which aggregates of size $k$ are formed in aggregative collisions of particles  $i$ and $j$ (the factor $\frac12$ avoids double counting). The second and third terms give the rates at which the particles of size $k$ disappear in collisions with other particles of any size $i$, due to aggregation and fragmentation, respectively. The fourth and fifth terms account for production of particles of size $k$ due to disruption of larger particles. Here $x_k(i)$ is the total number of debris of size $k$, produced in the
disruption of a projectile of size $i$. We have analyzed two models for the distribution of debris $x_k(i)$. One is the complete fragmentation model, $x_k(i)  = i \delta_{1k}$, when both colliding particles disintegrate into monomers; another is a power-law fragmentation model, when the distribution of debris sizes obeys a power-law, $x_k(i) \sim B(i) k^{-\alpha}$, in agreement with experimental  observations, see e.g.~\cite{Astrom2006,Guettler2010}; the impact of collisions with erosion is also analysed.

\subsection{Decomposition into monomers}
\label{secmon}
In the case of complete fragmentation,  $x_k(i)  = i \delta_{1k}$, the general kinetic  equations (\ref{kinpow}) become
\begin{eqnarray}
\label{eqmon}
\frac{dn_k}{dt} &=& \frac12 \sum_{i+j=k}  C_{ij} n_in_j - \sum_{i\geq 1} \left(C_{ik}+A_{ik}\right) n_in_k\\
\label{eqmon_n1}
\frac{d n_1}{dt} &=&   - n_1\sum_{j\geq 1}C_{1j}n_j + n_1\sum_{j\geq 2}jA_{1j}n_j\\
&+&\frac{1}{2} \sum_{i,j\geq 2}A_{ij}(i+j) n_in_j   \,. \nonumber
\end{eqnarray}
Mathematically similar equations modeling a physically different setting
(e.g., fragmentation was assumed to be spontaneous and collisional) have been analyzed in the context of rain drop  formation~\cite{Srivastava1982}.\\

\noindent \emph{{Constant rate coefficients.}}
The case of constant $C_{ij}=C_0$ and $A_{ij} =\lambda C_0$ can be treated analytically, providing useful insight into the general structure of solutions of Eqs.~(\ref{eqmon})--(\ref{eqmon_n1}), explicitly showing the emergence of the steady state. The constant $\lambda$ here characterizes the relative frequency of disruptive and aggregative collisions. Without loss of generality we set $C_0=1$. Solving the governing equations for mono-disperse initial conditions, $n_k(t=0) = \delta_{k,1}$, one finds
\begin{equation}
\label{n1t-sol} n_1(t) = \lambda_1\left[1+\lambda^{-1}\left(\lambda_2^{-1}e^{\lambda t}- \lambda^{-1}/2
\right)^{-\lambda_2/\lambda_1} \right]\, ,
\end{equation}
where $\lambda_1=\lambda/(1+\lambda)$ and  $\lambda_2= 2\lambda/(1+2\lambda)$. Utilizing the recursive nature of Eqs.~(\ref{eqmon}), one can determine $n_k(t)$ for $k > 1$. The system demonstrates a relaxation behavior: After a relaxation time that scales as $\lambda^{-1}$, the system approaches to a steady state with $n_1=\lambda_1$, the other concentrations satisfying
\begin{equation}
\label{eq:nk_cons_stead}
0=\frac12 \sum_{i+j=k} n_i n_j - (1+\lambda)n_k N.
\end{equation}
Here $N=\lambda_2$ is the steady-state value of the total number density of aggregates, $N=\sum_{k\geq 1} n_k$. We solve (\ref{eq:nk_cons_stead}) using the generation function technique to yield
\begin{equation}
\label{nk-sol}
n_k=\frac{N}{\sqrt{4\pi}} \left(1+\lambda\right)\left[\frac{2n_1}{(1+\lambda)N}\right]^k\,
\frac{\Gamma(k-\frac{1}{2})}{\Gamma(k+1)} \,.
\end{equation}
Now we assume that disruptive collisions in rings are considerably less frequent than aggregative ones, so that $\lambda \ll 1$ (this assumption leads to results that are consistent with observations);  moreover, $k \gg 1$ for most of the ring particles. Using the steady-state values, $n_1=\lambda_1$ and $N=\lambda_2$, one can rewrite Eq.~(\ref{nk-sol}) for $k \gg 1$ as
\begin{equation}
\label{nk-asymp-simple}
n_k=\frac{\lambda}{\sqrt{\pi}}\,e^{-\lambda^2 k}\,k^{-3/2} \,.
\end{equation}
Thus for $k<\lambda^{-2}$, the mass distribution exhibits power-law behavior,  $n_k \sim k^{-3/2}$, with an exponential cutoff for larger $k$.\\

\noindent\emph{{Size-dependent rate coefficients. }}
For a more realistic description, one must take into account the dependence of the rate coefficients on the aggregate size (Eqs.\ (\ref{eq:KinCoef})). Here we present the results for two basic limiting cases that reflect the most prominent features of the system:

1.\ The first case corresponds to energy equipartition,
$\left< E_k \right> =\frac12 m_k\left< {\bf v}_k^2 \right>={\rm const}$, which implies that the energy of random motion is equally distributed among  all species, like in molecular gases. In systems of dissipatively colliding particles, like planetary rings, this is usually not fulfilled, the smaller particles being colder than suggested by equipartition \cite{salo1992a,bodrova2014}. We also assume that the threshold energies of aggregation and fragmentation are constant: $E_{\rm agg} ={\rm const}$ and $E_{\rm frag}={\rm const}$; the latter quantities may be regarded as effective average values for all collisions. Then, as it follows from
Eqs.~(\ref{eq:KinCoef}), we have  $\lambda=A_{ij}/C_{ij}=\rm const$, and the kinetic coefficients read
\begin{equation}
\label{Cbal}
C_{ij}=C_0 \left( i^{1/3}+j^{1/3} \right)^2 \left( i^{-1}+j^{-1} \right)^{1/2} \,,
\end{equation}
where $C_0=\rm const$, so that the $C_{ij}$ are homogeneous functions of the masses of colliding particles\begin{equation}
\label{eq:C_hom}
C_{ai, aj} = a^{\varkappa} C_{ij}\,.
\end{equation}
The specific form (\ref{Cbal}) implies that the homogeneity degree is
$\varkappa =1/6$.

2. The second limiting case corresponds to equal velocity dispersion for \emph{all} species, $\left< {\bf v}_i^2 \right> = {\bf v}_0^2= {\rm const}$. In planetary rings the smaller particles \emph{do} have larger velocity dispersions than the larger ones but they are by far not as hot as equipartition would imply \cite{salo1992a}. Thus, this limiting case of equal velocity dispersions is closer to the situation in the rings.
For the dependence of the fragmentation threshold energy $E_{\rm frag}$ on the masses of colliding aggregates we employ the symmetric function $E_{{\rm frag}} = E_0\frac{ij}{i+j}$, which implies that  $E_{{\rm frag}}$ is proportional to the reduced mass of the colliding pair, $\mu_{ij} =m_1\frac{ij}{i+j}$. This yields $B_{ij} E_{\rm frag} =\left(\frac{3E_0}{2m_1 {\bf v}_0^2} \right) =\rm const$ and allows a simplified analysis. We assume that the aggregation threshold energy $E_{{\rm agg}}$ for all colliding pairs is large compared to the average kinetic energy of the relative motion of colliding pairs, $ \frac12 \mu_{ij} {\bf v}_0^2$ (our detailed analysis confirms this assumption, see the SI Appendix).
Then $\exp(-B_{ij} E_{\rm agg}) \ll  1$ and  Eqs.~(\ref{eq:KinCoef}) yield $C_{ij} = \nu_{ij}$. Therefore the ratio $A_{ij}/C_{ij} = \exp(-B_{ij} E_{\rm frag})$ is again constant. Thus  the relative frequency of disruptive and aggregative collisions is also characterized by the constant $\lambda = A_{ij}/C_{ij}$. The kinetic coefficients attain now the form
\begin{equation}
\label{Cjur}
C_{ij}=\tilde{C}_0 \left( i^{1/3}+j^{1/3} \right)^2,
\end{equation}
which is again a homogeneous function of $i$ and $j$ but with different homogeneity degree $\varkappa = 2/3$.

An important property of the kinetic equations, where the rate coefficients $C_{ij}$ and $A_{ij}= \lambda C_{ij}$ are homogeneous functions of  $i$ and $j$, is that these equations possess a scaling solution for $i,j \gg 1$. The latter is determined by the homogeneity degree $\varkappa$ and is practically insensitive to the detailed form of these coefficients \cite{Leyvraz2003,krapbook}. We use this property and replace the original rate coefficients (\ref{Cbal}) and (\ref{Cjur}) by the generalized product kernel
\begin{equation}
\label{Csimple}
C_{ij}=\widehat{C}_0 (ij)^{ \mu} \qquad \qquad \widehat{C}_0=\rm const
\end{equation}
For this kernel, the homogeneity degree is $\varkappa=2\mu$. To match it with the homogeneity degree of (\ref{Cbal}) and (\ref{Cjur}) we choose $\mu =1/12$ for the first limiting case and $\mu =1/3$ for the second. The advantage of the product kernel (\ref{Csimple}) is the existence on an analytic solution for the steady state distribution. Indeed, with the homogeneous coefficients (\ref{Csimple}) the steady-state version of
Eqs.~(\ref{eqmon}) reads,
\begin{equation}
\label{mk} 0 = \frac12 \sum_{i+j=k} l_i l_j - (1+\lambda)l_k L
\end{equation}
where we have used the shorthand notations
\begin{equation}
l_k = k^\mu n_k, \qquad \qquad L = \sum_{k\geq 1} l_k \,.
\end{equation}
With the substitute, $n_k \to l_k$ and $N \to L$, the system of equations (\ref{mk}) is mathematically identical to the system of equations with a constant kernel (\ref{eq:nk_cons_stead}), so that the steady-state solution reads,
\begin{equation}
\label{nk0} n_k=\frac{L}{2\sqrt{\pi}}\,e^{-\lambda^2 k}\,k^{-3/2-\mu} \,,
\end{equation}
again a  power-law dependence with exponential cutoff.

Our analytical findings  are confirmed by simulations. In Fig.~1, 
the results of a direct numerical solution of the system of rate equations (\ref{eqmon})--(\ref{eqmon_n1}) are shown for both limiting kernels, (\ref{Cbal}) and (\ref{Cjur}), together with their simplified counterparts (\ref{Csimple}). The stationary distributions for the systems with the complete kinetic coefficients (\ref{Cbal})--(\ref{Cjur}) have exactly the same slope as the systems with the simplified kernel (\ref{Csimple}) of the same degree of homogeneity and hence \emph{quantitatively agree} with the theoretical prediction (\ref{nk0}). Moreover, the numerical solutions demonstrate an exponential cutoff for large $k$, in agreement with the theoretical predictions.

Kernels (\ref{Cbal}) and (\ref{Cjur}) with homogeneity degree $\varkappa =1/6$ and $\varkappa =2/3$ correspond to two limiting cases of the size dependence of the average kinetic energy $\left< E_k \right> = \frac12 m_k \left< {\bf v}_k^2 \right>\sim k^{\beta}$. Namely,  $\beta =0$ corresponds to  $\varkappa =1/6$ and $\beta =1$ to $\varkappa =2/3$. Physically, we expect that $\beta$ is constrained within the interval $0 \leq \beta \leq 1$. Indeed, negative $\beta$ would imply vanishing velocity dispersion for very large particles, which is only possible for the unrealistic condition of the collision-free motion. The condition $\beta >1$ is unrealistic as well, since it would imply that the energy in the random motion, pumped by particle collisions, increases with particle mass faster than linearly; there is no evidence for such processes. We conclude that $\beta$ must be limited within the interval $[0,1]$, and therefore $\mu=\varkappa/2$ varies in the interval $1/12 \leq \mu  \leq 1/3$.

\subsection{Power-law collisional decomposition and erosion}
\label{secpow}
Although the fragmentation model with a complete  decomposition into monomers allows a simple analytical treatment, a more realistic model implies power-law size distribution of debris \cite{Astrom2006,Guettler2010}; moreover, collisions with erosion also take place \cite{Guettler2010,BlumErosion2011}. \\

\noindent\emph{Power-law decomposition.}
Experiments \cite{Astrom2006,Guettler2010} show that the number of debris particles of size $k$ produced in
the fragmentation of a particle of size $i$ scales as $x_k(i) \sim B(i)/k^{\alpha}$. If the distribution $x_k(i)$ of the debris size is steep enough,  the emerging steady-state particle distribution should be close to that for complete fragmentation into monomers. A scaling analysis, outlined below  confirms this expectation,
 provided that $\alpha >2$; moreover, in this case $B(i)=i$ (see SI Appendix).

Substituting the debris size distribution $x_k(i) \sim i /k^{\alpha}$ into the basic kinetic equations (\ref{kinpow}), we notice that the equation for
the monomer production rate coincides with Eq.~(\ref{eqmon_n1}), up to a factor in the coefficients $A_{ij}$. At the same time, the general equations (\ref{kinpow}) for $n_k$ have the same terms as Eqs.~(\ref{eqmon}) for complete decomposition, but with two extra terms -- the forth and fifth terms in Eqs.~(\ref{kinpow}). These terms describe an additional gain of particles of size $k$ due to decomposition of larger aggregates. Assuming that the steady-state distribution has the same form as for monomer decomposition, $n_k \sim k^{-\gamma} e^{-\nu k}$, one can estimate (up to a factor) these extra terms for the homogeneous kinetic coefficients, $A_{ij} =\lambda C_{ij} \sim (ij)^{\mu}$ [see Eq.~(\ref{Csimple})]. One gets
\begin{eqnarray}
\label{eq:sum1_pl}
&& \sum_{i=1}^k\sum_{j=k+1}^{\infty} A_{ij}n_in_j B(j)k^{-\alpha} \sim k^{-\alpha} \\
\label{eq:sum2_pl}
&&\sum_{i,j \geq k}  A_{ij}n_in_j \left[B(i)+B(j) \right] k^{-\alpha} \sim
k^{\mu-\gamma +1-\alpha}\,.
\end{eqnarray}
Here we also require that $\nu k \ll 1$, which is the region where the size distribution exhibits a
power-law behavior. The above terms are to be compared with the other three terms in Eqs.~(\ref{kinpow}) or Eqs.~(\ref{eqmon}), which are the same for monomer and power-law decomposition:
\begin{equation}
\label{eq:sum3_pl}
\sum_{i+j=k}  C_{ij}n_in_j  \sim \sum_{i \geq 1}   C_{ik}n_in_k  \sim \sum_{i
\geq 1}   A_{ik}n_in_k  \sim k^{\mu -\gamma} \,.
\end{equation}
If the additional terms (\ref{eq:sum1_pl}) and  (\ref{eq:sum2_pl}) were  negligible, as compared to the terms (\ref{eq:sum3_pl}) that arise for both models, the emergent steady-state size distributions would be the same.  For $k \gg 1$,  one can neglect (\ref{eq:sum1_pl}) and (\ref{eq:sum2_pl}) compared to (\ref{eq:sum3_pl}) if $\alpha > \gamma -\mu$ and $ \alpha > 1$. Taking into account that the equations for the monomers for the two
models coincide when $ \alpha > 2$, we arrive at the following criterion for universality of the steady-state distribution: $\alpha > {\rm max} \left\{\gamma -\mu, 2 \right\}$. In the case of complete decomposition into monomers we have $\gamma =\mu +3/2$ [see (\ref{nk0})]. Hence the above criterion becomes $\alpha > 2$. In other words, if  $\alpha > 2 $, the model of complete decomposition into monomers yields the same steady-state size distribution as the model with \emph{any} power-law distribution of debris. \\
\\
\noindent\emph{Collisions with erosion.}
In collisions with erosion only a small fraction of a particle mass is chipped off~\cite{Guettler2010,BlumErosion2011,KrapivskyBenNaim2003}. Here we consider a simplified model of such collisions: It takes place when the relative kinetic energy exceeds the threshold energy $E_{\rm eros}$, which is smaller than the fragmentation energy $E_{\rm frag}$. Also, we assume that the chipped-off piece always contains $l$ monomers. Following the same steps as before one can derive rate equations that describe both disruptive and erosive collision. For instance, for complete decomposition into monomers the equation for $n_k$ with $k \geq l+2$ acquires two additional terms
$$
\lambda_e \sum_{i=1}^{\infty} C_{i\, k+l}n_i n_{k+l} -\lambda_e 
\sum_{i=1}^{\infty} C_{i k}n_i n_{k},
$$
with similar additional terms for $l+2 > k >1$ and for the monomer equation. Here $\lambda_e$ gives the ratio of the frequencies of aggregative and erosive collisions, which may be expressed in terms of $E_{\rm eros}$ (see SI Appendix). We assume that $\lambda_e$ is small and is of the same order of magnitude as $\lambda$. We also assume that $\lambda_e$ is constant and that $l\lambda_e \ll 1$. Then we can shown that for $k \gg 1$ the size distribution of aggregates $n_k$ has exactly the same form, Eq.~(\ref{nk0}), as for the case of purely disruptive collisions (see SI Appendix). \\

\noindent\emph{Universality of the steady-state distribution.}
The steady-state size-distribution of aggregates (\ref{nk0}) is generally \emph{universal}:~It is the same for all size distributions of debris, with a strong dominance of small fragments, independently of its functional form. Moreover, it may be shown analytically (see SI Appendix),  that the form (\ref{nk0}) of the distribution persists when collisions with erosion are involved. We checked this conclusion numerically, solving the kinetic equations (\ref{kinpow}) with a power-law, exponential size distribution of debris and for collisions with an erosion (Fig.~2). 
We find that the particle size distribution (\ref{nk0}) is indeed universal for steep distributions of debris size. Fig.~2 
also confirms the condition of universality of the distribution (\ref{nk0}), if $\alpha >2$ for power-law debris size-distributions.

A steep distribution of debris size, with strong domination of small fragments appears plausible from a physical point of view:~The aggregates are relatively loose objects, with a low average coordination number, that is, with a small number of bonds between neighboring constituents; it is easy to break such objects into small pieces. The erosion at collision also has no qualitative impact on the size distribution as long as the frequency of such collisions is significantly less than of the aggregative ones.

\subsection{Size distribution of the ring particles}
The distribution of the ring particles' radii, $F(R)$, is constrained by space- and earth-bound observations \cite{cuzzi2009}. To extract $F(R)$ we use the relation $R^3 =r_k^3 = kr_1^3$ (for spherical particles) in conjunction with  $n_k dk = F(R) dR$. We find that $n_k\sim k^{-3/2-\mu}\exp(-\lambda^2 k)$ implies
\begin{eqnarray}
\label{fR} F(R) &\sim&  R^{-q}e^{-(R/R_c)^3} \qquad q=5/2 + 3\mu \\
\label{lambdaR} R_c &=& r_1/\lambda^{2/3} \,.
\end{eqnarray}
Thus for $ R \ll R_c$ the distribution is algebraic with exponent $q = 5/2 + 3 \mu$, and the crossover to exponential behavior occurs at $R \sim R_c$.

We have shown that the exponent $\mu$ can vary within the interval $1/12 \le \mu \le 1/3$, and hence the exponent $q$ for the size distribution varies in the range $2.75 \le q \le 3.5$. This is in excellent agreement with observations, where values for $q$ in the range from $2.70$ to $3.11$ were reported~\cite{zebker1985,french2000b}. Fitting the theory to the particle size distribution of Saturn's A ring inferred from data obtained by the Voyager Radio Science Subsystem (RSS) during a radio occultation of the spacecraft by Saturn's A ring \cite{zebker1985}, we find $R_c = 5.5$m (Fig.~3). 
For $r_1$ in the plausible range from $1$cm to $10$cm \cite{cuzzi2009} we get (Eq.\ \ref{lambdaR}) $\lambda$ on the order of $10^{-4}$ to $10^{-3}$, which is the ratio of the frequencies of disruptive and coagulating collisions. It is also possible to estimate characteristic energies and the strength of the aggregates. Using the plausible range for random velocity,  $v_0=0.01-0.1{\rm cm/s}$~\cite{cuzzi2009}, we obtain values that agree with the laboratory measurements (see SI Appendix).

\section{Conclusion and Outlook}
We have developed a kinetic model for the particle size distribution in a dense planetary ring and showed that the steady-state distribution emerges from the dynamic balance between aggregation and fragmentation processes. The model quantitatively explains properties of the particle size distribution of Saturn's rings inferred from observations. It naturally leads to a power-law size distribution with an exponential cutoff (Eq.~\ref{fR}).
Interestingly, the exponent $q=2.5+3\mu$ is universal, for a specific class of models we have investigated in detail. That is, $q$ does not depend on details of the collisional fragmentation mechanism, provided the size distribution of debris, emerging in an impact, is steep enough; collisions with erosion do not alter $q$ as well. The exponent $q$ is a sum of two parts: The main part, $2.5$, corresponds to the "basic" case when the collision frequency does not depend on particle size ($\mu=0$); such slope is generic for a steady size-distribution, stemming from the aggregation-fragmentation balance in \emph{binary} collisions. The additional part, $3\mu$, varying in the interval $0.25 \leq 3\mu \leq 1$, characterizes size dependence of the collision  frequency. The latter is determined by the particles' diameters and the mean square velocities $\left< {\bf v}_k^2 \right>$ of  their random motion. We have obtained analytical solutions for the limiting cases of  energy equipartition, $\frac12 m_k \left< {\bf v}_k^2 \right>={\rm const}$, ($3\mu=0.25$) and of equal velocity dispersion for all species $\left< {\bf v}_k^2 \right>={\rm const}$, ($3\mu=1$). These give the limiting slopes of $q=2.75$ and $q=3.5$. Physically, we expect that an intermediate dependence between these two limiting cases may follow from a better understanding of the behaviors of threshold energies. This would imply a power-law size distribution with exponent in the range $2.75 \leq q \leq 3.5$.

Observed variations of spectral properties of ring particles \cite{NICHOLSON:2008ex,Filacchione:2012ff} may indicate differences in the surface properties, and thus, in their elasticity and sticking efficiency. This implies differences in the velocity dispersion $\left< {\bf v}_k^2 \right>$ and its dependence on $k$, resulting in different values of the exponent $q$. Moreover, variations in particle sizes among different parts of Saturn's ring system have been inferred from Cassini data \cite{cuzzi2009,2013AGUFM.P21E..01C}. For our model, a different average particle size, or monomer size, implies different values of $E_{\rm frag}$ and $E_{\rm agg}$ as well as different values of the upper cutoff radius $R_c$. The model gives $R_c$ in terms of the primary grain radius and the ratio of the disruptive  and  aggregative collisions, Eq.~(\ref{lambdaR}). Since this ratio increases with increasing kinetic energy of particles' random motion, the cutoff radius is expected to be smaller for rings with larger velocity dispersion.

Our results essentially depend on three basic assumptions:~(i) ring particles are aggregates comprised from primary grains which are kept together by adhesive (or gravitational) forces; (ii) the aggregate sizes change due to \emph{binary} collisions, which are aggregative, bouncing, or disruptive (including collisions with erosion); (iii) the collision rates and  type of impacts are determined by sizes and velocities of colliding particles. We wish to stress that the power-law distribution with a cutoff is a \emph{direct mathematical consequence} of the above assumptions only, that is, there is no need to suppose a power-law distribution and search for an additional mechanism for a cutoff as in previous semi-quantitative approaches \cite{longaretti1989}.

The agreement between observations and predictions of our model for the size distribution indicates that dense planetary rings are indeed mainly composed of aggregates similar to the Dynamic Ephemeral Bodies suggested three decades ago~\cite{davis1984,weidenschilling1984}. This means that the (ice) aggregates constituting the cosmic disks permanently change their mass due to collision-caused aggregation  and fragmentation, while their distribution of sizes remains stationary. Thus, our results provide another (quantitative) proof that the particle size distribution of Saturn's rings is not primordial. The same size distributions are expected for other collision dominated rings, as the rings of Uranus \cite{elliot1984a,elliot1984b}, Chariklo \cite{BragaRibas2014,2014A&A...568A..79D} and Chiron \cite{2015A&A...576A..18O,Ruprecht:2015ho}, and possibly around extrasolar objects \cite{ohta2009,2012AJ....143...72M,2015MNRAS.446..411K}.

The predictive power of the kinetic model further emphasizes the role of adhesive contact forces between constituents which dominate for aggregate sizes up to the observed cutoff radius of $R_c \sim 5-10 \, {\rm m}$. The model does not describe the largest constituents in the rings, with sizes beyond $R_c$. These are the propeller-moonlets in the A and B rings of Saturn, which may be remnants of a catastrophic disruption~\cite{Sremcevic2007,esposito2006}. We also do not discuss the nature of the smallest constituents, that is, of the primary grains. These particles are probably themselves comprised of still smaller entities and correspond to the least-size free particles observed in the rings \cite{icarus2012}.

Recently, cohesion was studied for dense planetary rings in terms of N-Body simulations \cite{Perrine2011,Perrine2012}. This model is similar to ours in that the authors use critical velocities for merging and fragmentation while we use threshold energies $E_{\rm agg}$ and $E_{\rm frag}$; both criteria are based on the cohesion model. In these simulations a power-law distribution for the aggregates size $F(R)\sim R^{-q}$ was obtained with slopes $2.75 \leq q \leq 3$ for reasonable values of the cohesive parameter, consistent with our theoretical result. Moreover, the critical velocities for merging and fragmentation differ in most of the simulations by a factor of two, which is in reasonable agreement with our model, where we estimated $E_{\rm frag} $ to be roughly twice $E_{\rm agg}$ (SI Appendix).  However, the simulations cannot resolve an exponential cutoff of $F(R)$, due to the small number of large aggregates, but the dependence of the largest aggregate size is inferred for different cohesion parameters and critical velocities.

\section{Materials and Methods}
\subsection{Boltzmann equation}
The general equations (\ref{kinpow}) for the concentrations  $n_k$ have been derived from the Boltzmann kinetic equation. Here we consider a simplified case of a force-free and spatially uniform system. It is possible to take into account the effects of non-homogeneity, as it is observed in self-gravity wakes, and gravitational interactions between particles. These, however, do not alter the form of resulting rate equations (\ref{kinpow}), which may be then formulated for the space-averaged values (see SI Appendix).

Let $f_i \equiv f(m_i,{\bf v}_i,t)$ be the mass-velocity distribution function which gives density  of particles of mass $m_i$ with the velocity ${\bf v}_i$ at time $t$. In the homogeneous setting, the distribution function evolves according to the Boltzmann equation,
\begin{equation}
\label{eq:BEgen} \frac{\partial}{\partial t} f_k\left({\bf v}_k, t \right) = I^{\rm b}_{k}+ I^{\rm
heat}_k+I^{\rm agg}_{k}+I^{\rm frag}_{k}\, ,
\end{equation}
where the right-hand side accounts for particles collisions.  The first term $I^{\rm b}_{k}$ accounts for bouncing collisions of particles (see e.g.~\cite{brillOUP}),  the second term $I^{\rm heat}_k$ describes the viscous heating caused  by the Keplerian shear (see e.g. \cite{schmidt2009}); the terms $I^{\rm agg}_{k}$  and $I^{\rm frag}_{k}$ account,  respectively, for the aggregative and disruptive impacts (explicit expressions for these terms are given in SI Appendix). To derive  the rate equations (\ref{kinpow}) for the concentrations of the species, $n_k(t)
= \int d {\bf v}_k f_k ({\bf v}_k,t)$, one needs to integrate Eq.~(\ref{eq:BEgen}) over  ${\bf v}_k$. Assuming that all species have a Maxwell velocity distribution function with average velocity $\left< {\bf v}_k \right>=0$ and velocity dispersion $\left< {\bf v}_k^2  \right>$ we obtain the rate equations (\ref{kinpow}) and the rate coefficients (\ref{eq:KinCoef}) (see SI Appendix for the detail).

\subsection{Generating Function Techniques}
To solve Eqs.~(\ref{eq:nk_cons_stead}) we use the generating function $\mathcal{N}(z)=\sum_{k\geq 1} n_k\,z^k$ which allows us to convert these equations into the single algebraic  equation
\begin{equation}
\label{Nz-eq} \mathcal{N}(z)^2 -2(1+\lambda)N\mathcal{N}(z) + 2(1+\lambda)Nn_1z=0\,.
\end{equation}
Its solution reads
\begin{equation}
\label{Nz-sol} \mathcal{N}(z) = \left(1+\lambda\right)N\left[1 - \sqrt{1-
\frac{2n_1}{(1+\lambda)N}\,z}\,\right]\,.
\end{equation}
Expanding $\mathcal{N}(z)$ we arrive at the distribution (\ref{nk-sol}).

\subsection{Efficient numerical algorithm}
  $\,$ The numerical solution of Smoluchowski-type equations (\ref{kinpow}) is challenging as one has to solve infinitely many coupled non-linear equations. We developed an efficient and fast numerical algorithm dealing with a large number of such equations. The application of our algorithm requires the condition
\begin{equation}
\label{eq:cond}
\sum_{i=1}^{k}C_{i,k+1}n_{i} \gg \sum_{i=k+1}^{N}C_{i,k+1}n_{i}
\end{equation}
is obeyed for $k \gg 1$ and $N \gg 1$, where $n_i$ are the steady-state concentrations. For the case of interest this condition is  fulfilled. Our algorithm first solves a relatively small set ($\sim 100$) of equations using the standard technique and then obtains  other concentrations of  much larger set ($\sim 10,000$) using an iterative procedure (see SI Appendix).

\begin{acknowledgments}
We thank Larry Esposito, Heikki Salo, Martin Sei{\ss}, and Miodrag Srem\v cevi\'c for fruitful discussions. Numerical calculations were performed using Chebyshev supercomputer of Moscow State University. This work was supported by Deutsches Zentrum f\"ur Luft und Raumfahrt, Deutsche Forschungsgemeinschaft, Russian Foundation for Basic Research (RFBR, project 12-02-31351). The authors also acknowledge the partial support through the EU IRSES DCP-PhysBio N269139 project.
\end{acknowledgments}


\begin{figure}
\label{fig:difkern}
\centerline{\includegraphics[width=0.38\textwidth,angle=270]{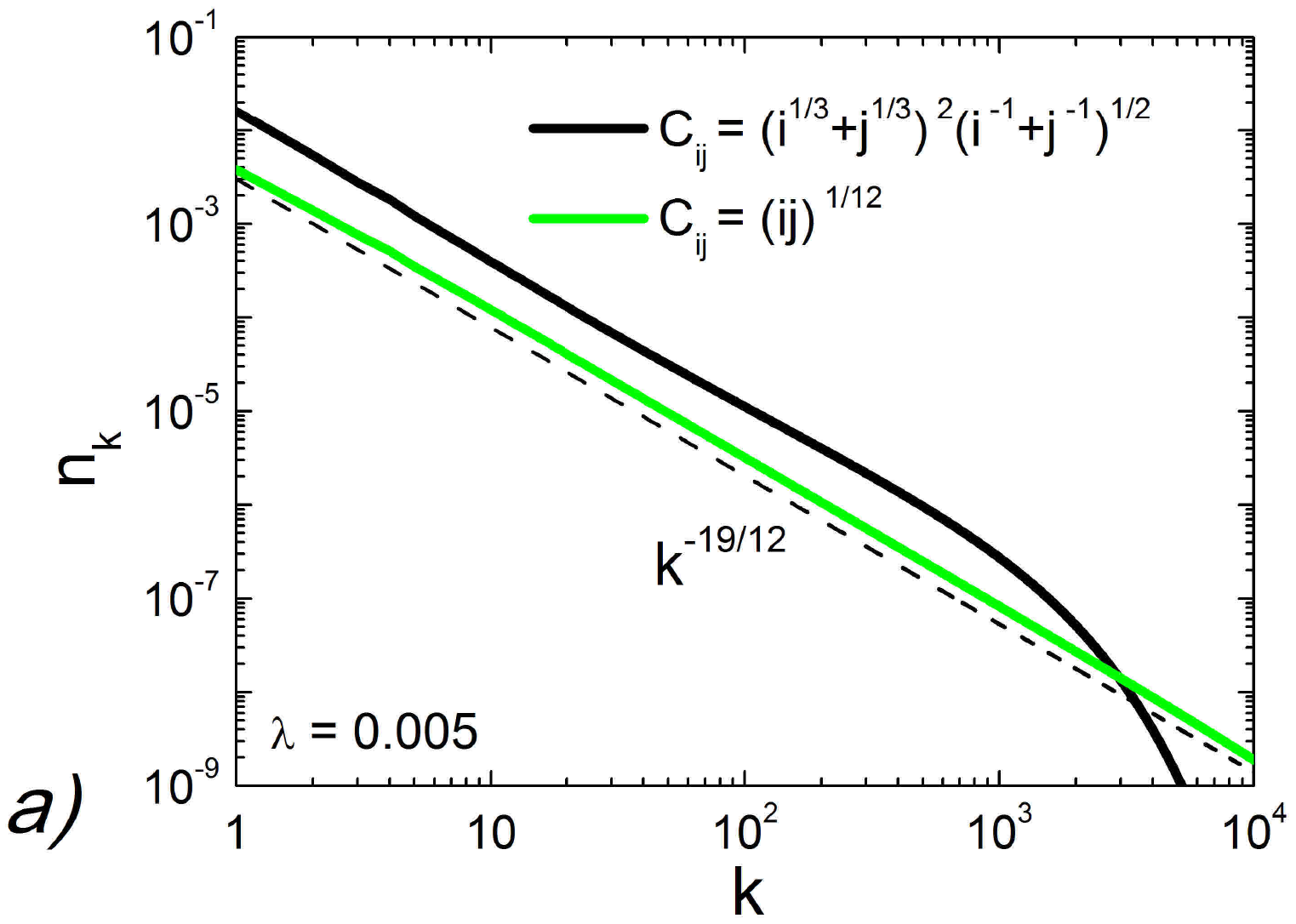}}
\centerline{\includegraphics[width=0.38\textwidth,angle=270]{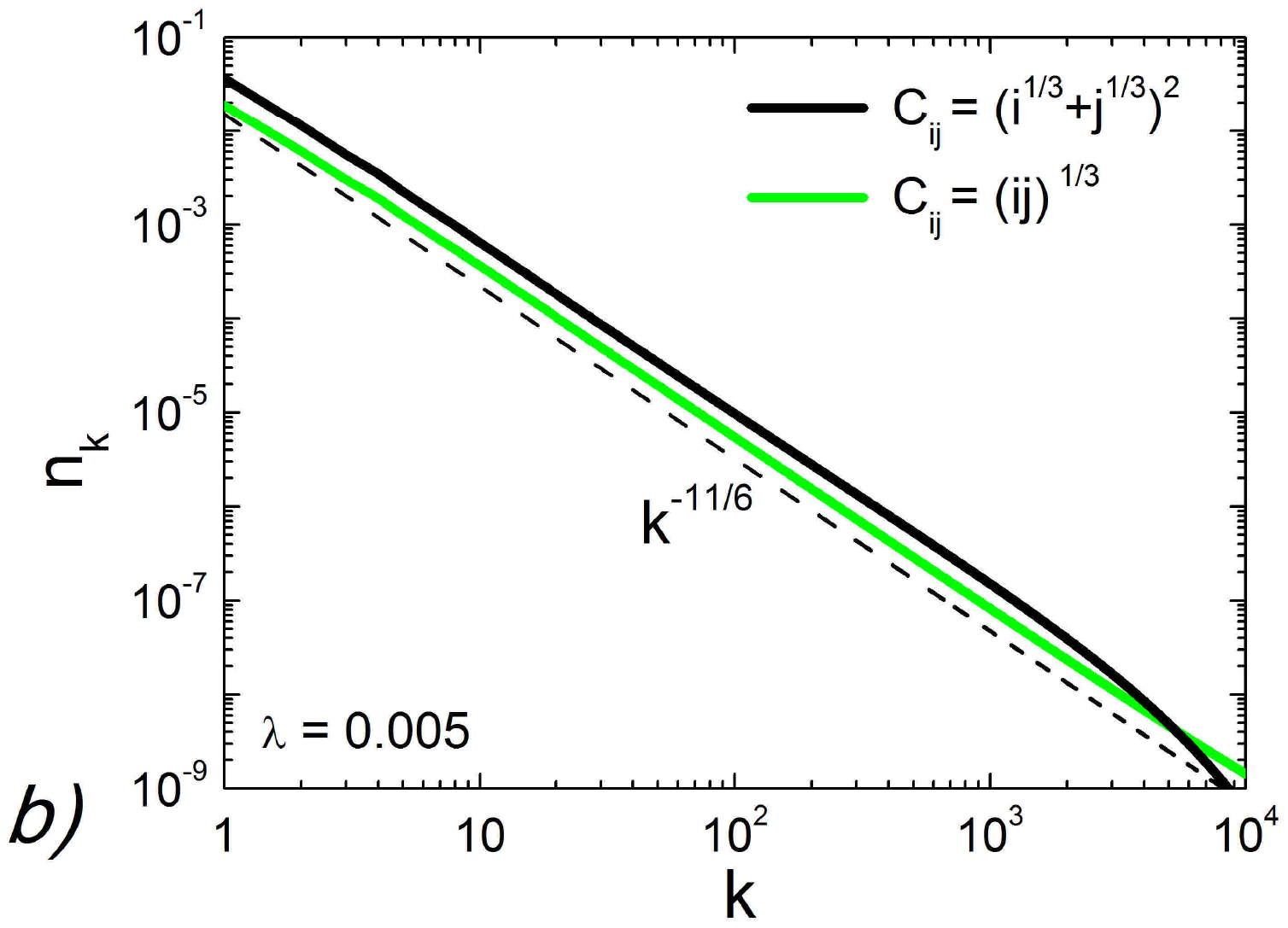}}
\caption{ {\bf Particle size distribution in the case of complete decomposition into monomers.} (a) The limiting case of energy equipartition $\left< E_k  \right> ={\rm const}$ for all species. The solid black line corresponds to the system with complete kinetic coefficients (\ref{Cbal}), while the solid green line corresponds to the simplified coefficients (\ref{Csimple}) with the same degree of homogeneity, $2 \mu =1/6$. The dashed line has slope $n_k \propto k^{-19/12}$,
predicted by the theory. (b) The limiting case of equal velocity dispersion for all species, $\left< {\bf v}_k^2 \right> ={\rm const}$. The solid black line refers to the system with the complete kinetic coefficients (\ref{Cjur}), while the solid green line corresponds to the simplified coefficients (\ref{Csimple}) with the same degree of homogeneity, $2 \mu =2/3$. The dashed line shows the theoretical dependence,  $n_k \propto k^{-11/6}$. In all cases the power-law distribution turns into an exponential decay for large sizes.}
\end{figure}

\begin{figure}
\centerline{\includegraphics[width=0.6\textwidth,angle=0]{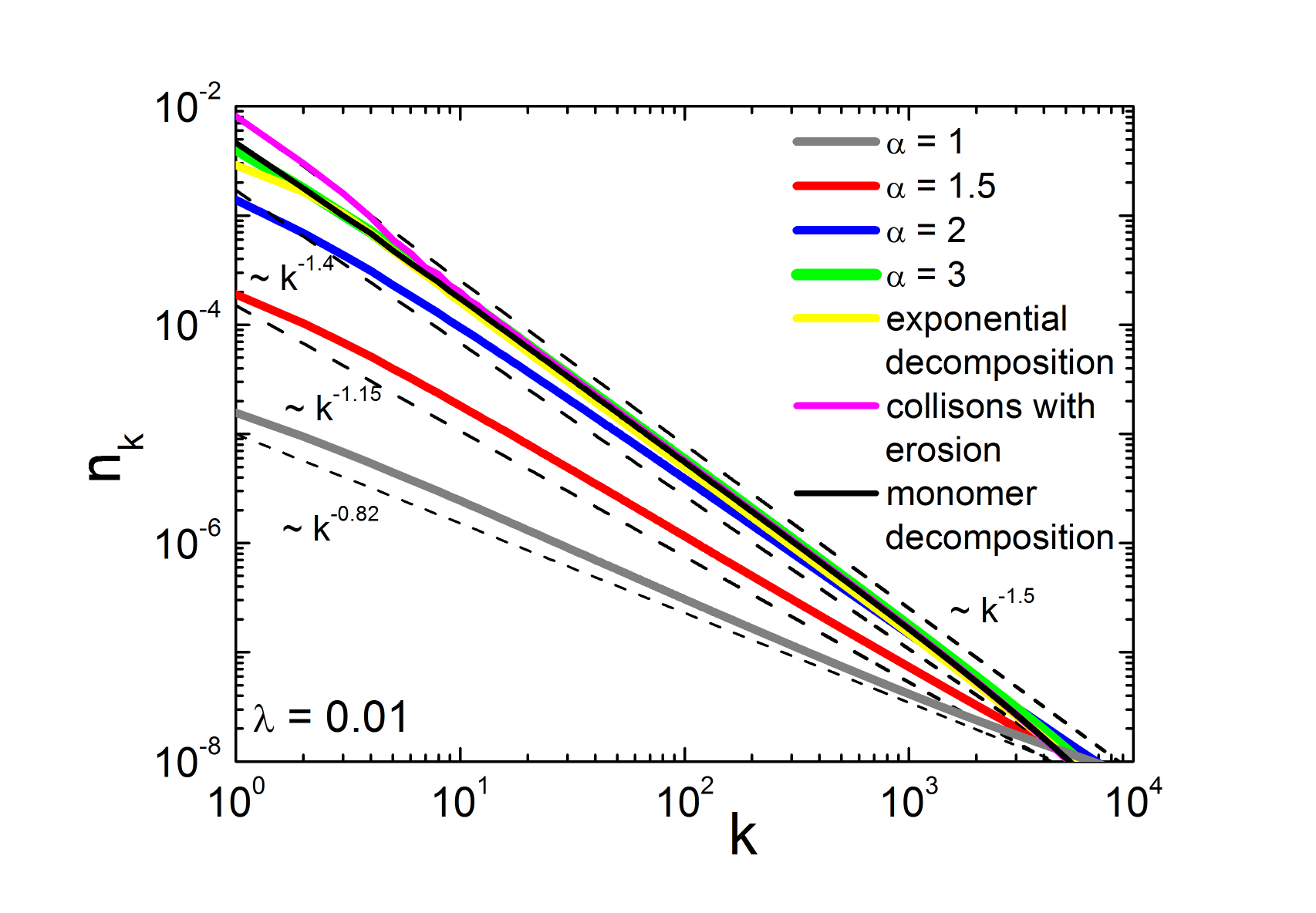}}
\caption{ {\bf Particle size distribution in the case of power-law decomposition}. The black solid line depicts the particle size distribution for the case of complete decomposition.   The other solid lines show the emergent steady-state distributions for the following size distributions of debris:~the power-law  distribution, $x_k(i) \sim k^{-\alpha}$, with $\alpha = 1$ (gray), $\alpha = 1.5$ (red), $\alpha= 2$ (blue), $\alpha=3$ (green), exponential distribution $x_k(i) \sim \exp\left(-k\right)$ (yellow) and collisions with an erosion, $\lambda_e=0.05$, $l=3$ (violet).  The dashed lines indicate the corresponding  fit $n_k \sim k^{-\gamma}$. Note that for steep size distributions of debris (power law with $\alpha > 2$, exponential distribution) all slopes coincide with the one for the case of complete decomposition into monomers. All curves correspond to the case of constant kinetic coefficients with  $\lambda=C_{ij}/A_{ij}=0.01$.} \label{Gpow}
\end{figure}

\begin{figure}
\centerline{\includegraphics[width=0.425\textwidth,angle=90]{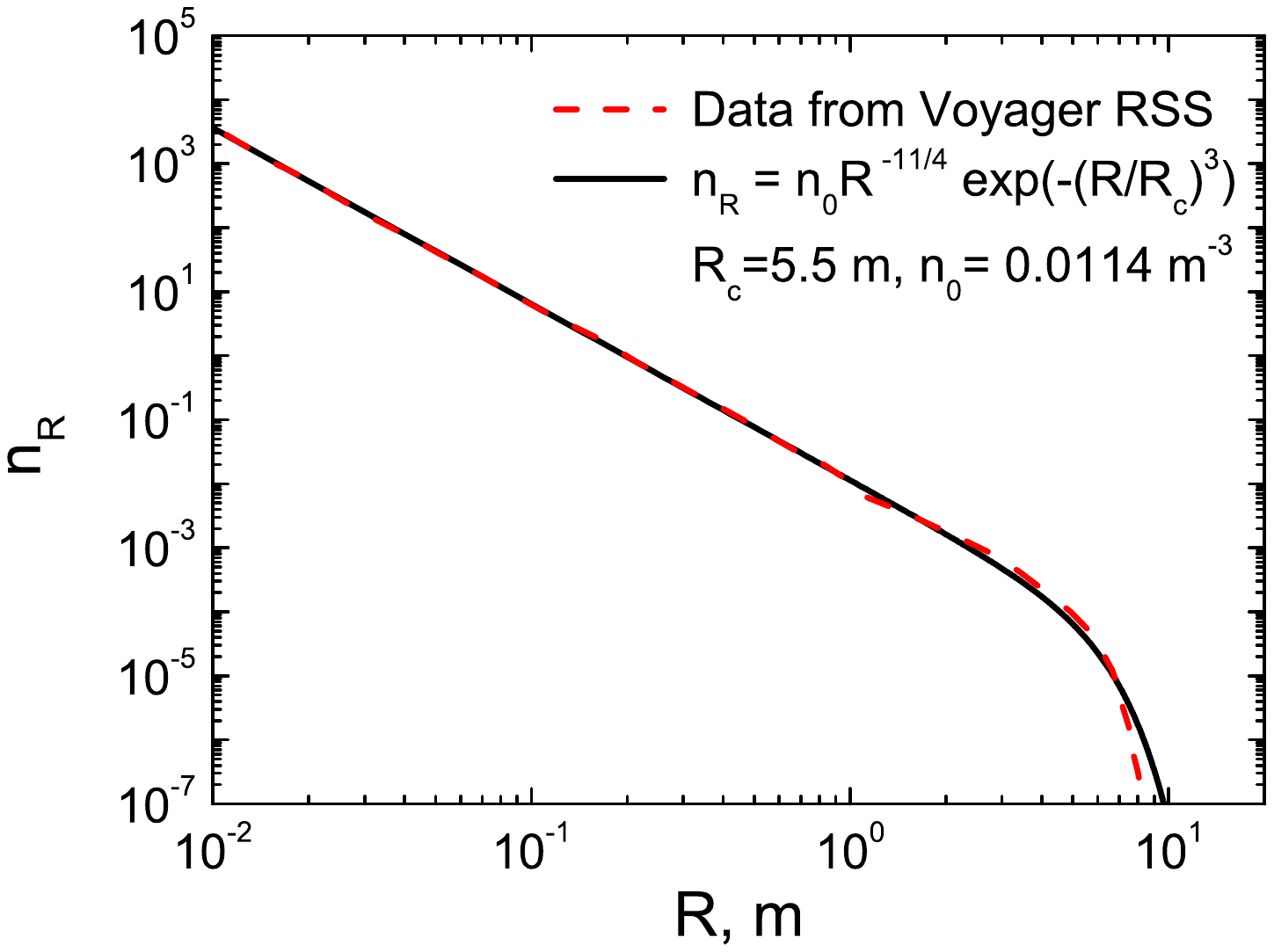}}
\caption{
 {\bf Particle size distribution for Saturn's A ring}. The dashed line represents the particle size
 distribution of Saturn's A ring inferred from data obtained by the Voyager Radio
 Science Subsystem (RSS) during a radio occultation of the spacecraft by the
 rings~\cite{zebker1985}. A fit of  the theoretical curve, Eq.(\ref{fR}), is
 shown as a solid line, giving a cutoff radius of $R_c = 5.5\,  {\rm m}$}\label{Gzebker}
\end{figure}

\end{article}

\end{document}


\title{Supporting Information}
\author{Nikolai Brilliantov\affil{1}{University of Leicester, Leicester, United Kingdom},
P. L. Krapivsky\affil{2}{Department of Physics, Boston University, Boston, USA},
Anna Bodrova\affil{3}{Moscow State University, Moscow, Russia}\affil{4}{University of Potsdam, Potsdam, Germany},
Frank Spahn\affil{4}{},
Hisao Hayakawa\affil{5}{Yukawa Institute for Theoretical Physics, Kyoto University, Kyoto, Japan},
Vladimir Stadnichuk\affil{3}{},
\and J\"urgen Schmidt\affil{6}{Astronomy and Space Physics, University of Oulu, Finland}\affil{4}{}}
\maketitle
\begin{article}

\section{Collision integrals}
The collision integral for aggregative collisions has the following form:
\begin{eqnarray}
\nonumber &&\!I^{\rm agg}_{k}(\vec{v}_k)  = \frac12 \! \sum_{i+j=k} \! \sigma_{ij}^2 \!\int \!d
\vec{v}_i\!\int \!d \vec{v}_j \!\int \! \!d\vec{e}\,\Theta\left(-\vec{v}_{ij} \cdot \vec{e} \, \right)
\left|\vec{v}_{ij} \cdot \vec{e} \,\right| \\    && \times  f_i\left(\vec{v}_i \right) f_j\left(\vec{v}_j
\right)\Theta\left(E_{\rm agg}-E_{ij} \right) \delta(m_k \vec{v}_k-m_i \vec{v}_i-m_j \vec{v}_j )
\nonumber \\
&&\!- \sum_{j}\sigma_{kj}^2 \int d \vec{v}_j \int d\vec{e} \, \Theta\left(-\vec{v}_{kj} \cdot \vec{e} \,
\right) \left|\vec{v}_{kj} \cdot \vec{e} \,\right|     \nonumber \\    &&\times f_k\left(\vec{v}_k \right)
f_j\left(\vec{v}_j \right) \Theta\left(E_{\rm agg}-E_{kj}\right) =I^{\rm agg, 1}_k - I^{\rm agg, 2}_k\, . \label{eq:Iagg}
\end{eqnarray}
The first sum on the right-hand side of Eq.~(\ref{eq:Iagg}) refers to collisions where an aggregate of
mass $k$ is formed from smaller aggregates of masses $i$ and $j$, while the second sum describes the
collisions of $k$-aggregates with all other particles. In the first sum $m_k=m_i+m_j$ and  $m_k \vec{v}_k =
m_i\vec{v}_i +m_j \vec{v}_j$ due to mass and momentum conservation. The rest of the notation is standard, see
e.g.~\cite{brillOUP}: $\sigma_{ij}^2 =r_1\left( i^{1/3}+j^{1/3} \right)^2$ quantifies the collision
cross-section and $\left|\vec{v}_{ij} \cdot \vec{e} \,\right|$ is the length of the collision cylinder, where
the unit vector $\vec{e}$ specifies the direction of the inter-center vector at the collision instant;
$\Theta\left(-\vec{v}_{ij} \cdot \vec{e} \, \right)$ selects only approaching particles. The factor
 $\Theta\left(E_{\rm agg}-E_{ij}  \right)$  in the integrands guarantees that the relative kinetic energy does not exceed $E_{\rm agg}$ to cause the aggregation. The
kinetic energy of the relative motion is $E_{ij} =\frac12 \mu_{ij} \vec{v}_{ij}^{\,2} $, with the
relative velocity $\vec{v}_{ij}=\vec{v}_i-\vec{v}_j$ and reduced mass
$\mu_{ij}=m_im_j/(m_i+m_j)$. The notations in the second sum on the right-hand side of Eq.~(\ref{eq:Iagg})
have the similar meaning.

For the collisions leading to fragmentation we have
\begin{eqnarray}\nonumber
 &&I^{\rm frag}_{k} (\vec{v}_k) =\frac12 \sum_{i,j\geq k+1} \sigma_{ij}^2 \int d
\vec{v}_j\int \!d \vec{v}_i \int d \vec{e} \, \Theta\left(-\vec{v}_{ij} \cdot \vec{e} \, \right)
   \nonumber \\
   &&\times \left|\vec{v}_{ij} \cdot \vec{e} \,\right| f_j\left(\vec{v}_j\right)f_i\left(\vec{v}_i \right)
   \Theta\left(E_{ij}^{\rm n} - E_{\rm frag} \right) \\ \nonumber
   &&\times   \left( q_{ki}  (\vec{v}_k, \, \vec{v}_i,\, \vec{v}_j )
   + q_{kj}  (\vec{v}_k, \, \vec{v}_i,\, \vec{v}_j ) \right) \\
&&+\sum_{i=1}^k\sum_{j\geq k+1} \sigma_{ij}^2 \int d \vec{v}_j\int \!d \vec{v}_i \int d \vec{e} \,
\Theta\left(-\vec{v}_{ij} \cdot \vec{e} \, \right)
   \nonumber \\ \nonumber
   &&\times \left|\vec{v}_{ij} \cdot \vec{e} \,\right|
   f_j\left(\vec{v}_j\right)f_i\left(\vec{v}_i \right)   \Theta\left(E_{ij}^{\rm n} - E_{\rm frag} \right) q_{kj}  \left(\vec{v}_k, \, \vec{v}_i,\, \vec{v}_j ) \right)    \\
   &&-\sum_{i }\left(1-  \delta_{k,1}\right)\sigma_{ki}^2 \int d \vec{v}_i \int d\vec{e} \,   \Theta\left(-\vec{v}_{ki} \cdot \vec{e} \, \right) \left|\vec{v}_{ki} \cdot \vec{e} \,\right|   \nonumber \\   &&\!\times f_k\left(\vec{v}_k\right)  f_i\left(\vec{v}_i \right)   \Theta\left(E_{ki}^{\rm n} - E_{\rm frag} \right)\!=\!I^{\rm frag,1}_{k}\!+\!I^{\rm frag,2}_{k}\!-\!I^{\rm frag,3}_{k}. \nonumber
  \label{eq:Ifrag}
\end{eqnarray}
where we define the kinetic energy of the relative {\it normal}  motion, $E_{ij}^{\rm n} =\frac12 \mu_{ij}
(\vec{v}_{ij} \cdot \vec{e} )^2 $, with $(\vec{v}_{ij} \cdot \vec{e} )$ being the normal relative velocity.
In contrast to the case of aggregation where both normal and tangential components must be small,
so that the total energy of the relative motion $E_{ij}$ matters, for fragmentation only the
relative normal motion is important: Only normal motion  causes a compression and the subsequent breakage of
particles' material. Hence the kinetic energy of the relative {\it normal} motion, $E_{ij}^{\rm n}$ must
exceed some threshold. The first sum in Eq.~(\ref{eq:Ifrag}) describes the collision of particles  $i>k$ and
$j>k$ with the relative kinetic energy of the normal motion above the fragmentation threshold $E_{\rm frag}$;
both particles give rise to fragments of size $k$. Further, $q_{ki} (\vec{v}_k, \, \vec{v}_i,\, \vec{v}_j )$ indicates
the number of debris of mass $m_k=m_1\, k$ with the velocity $\vec{v}_k$, when a  particle of mass $m_i=m_1\,
i$ disintegrates in a collision with a particle of mass $m_j=m_1\, j$, provided that the pre-collision velocities are $\vec{v}_i$ and $\vec{v}_j $. Obviously, $q_{ki}=0$ if $k \ge i$. The function
$q_{ki} (\vec{v}_k, \, \vec{v}_i,\, \vec{v}_j )$ depends on a particular collision model. The second sum
describes the process, when only one particle with $j>k$ (but not with $i<k$) gives rise to debris of size
$k$. Finally, the third term accounts for the breakage of particles of size $k > 1$ in collisions with all
other particles.

In the present study we focus on the evolution of particle densities $n_k$. Therefore: (i)~the
particular forms of two other collision integrals -- for bouncing collisions, $I_k^{\rm b}$, and for the one describing viscous heating, $I_k^{\rm heat}$,  are not important, since
these terms do not change densities of the species and (ii)~it is sufficient to use a more simple distribution, $x_k(i)$, defined as

\begin{equation}\label{eq:qx}
  \int q_{ki} (\vec{v}_k, \vec{v}_i,\vec{v}_j) d \vec{v}_k = x_k(i),
\end{equation}
where $x_k(i)$ gives the total number of fragments of size $k$ in all possible disruptive collisions of a particles of size $i>k$. In Eq.~(\ref{eq:qx}) we exploit the "mean-field" approximation, that is, we assume that the \emph{averaged} distribution $x_k(i)$ depends neither on the size of the colliding partner $j$, nor on the velocities of $\vec{v}_i$ and $\vec{v}_j$, provided a fragmentation occurs.

Note, that while $I_k^{\rm b}$ describes the loss of energy in dissipative collisions,
$I_k^{\rm heat}$ characterizes the energy input, so that the average kinetic energy of all species $\left< E_k \right>$, $k=1,2, \ldots$ is kept in a steady state.

\section{Maxwell approximation for the velocity distribution functions}

The interplay between aggregation and fragmentation results in a dynamically sustained mixture of particles of
different mass. Mixtures of dissipative particles, generally, have different velocity dispersion, or mean
kinetic energy (``granular temperature") of each species. The partial number density (concentration) $n_i$ and
the respective mean kinetic energy $\langle E_i \rangle $ of the species read, e.g.~\cite{GarzoDufty:1999},

\begin{equation}
\label{eq:nTiTdef}
 n_i  =  \int d\vec{v}_i f_i (\vec{v}_i) \, ,  \quad
n_i \langle E_i \rangle  =  \int \frac{ m_i \vec{v}_i^{\, 2}}{2} f_i(\vec{v}_i)  d \vec{v}_i \, .
\end{equation}
We assume that the distribution function $f_i(\vec{v}_i,t )$ may be written as
\cite{SpahnetalEPL:2004,BrilliantovSpahn2006,GarzoDufty:1999}
\begin{equation}
\label{eq:NDFpartial} f_i(\vec{v}_i, t) = \frac{n_i(t)}{v_{0, \, i}^3(t)} \phi_i (\vec{c}_i) \,, \qquad
\vec{c}_i \equiv \frac{\vec{v}_i}{v_{0,\,i}} \, ,
\end{equation}
where $v_{0, \,  i}^2(t) = 2  \langle E_i \rangle (t)/m_i$ is the thermal velocity and $ \phi (c_i) $ the
reduced distribution  function. For force-free granular mixtures~\cite{GarzohrenyaDuftyII} and interacting
particles (which suffer ballistic annihilation)~\cite{Trizac_anih} the reduced distribution function is
represented in the form of the Sonine polynomial expansion,
$$
\phi_i (\vec{c})=\phi_M(\vec{c})\left[1 + \sum_{k=1}^{\infty}a_k^{(i)} S_k(c^2) \right]
$$
Here $\phi_M(\vec{c})$ is the Maxwell distribution  function,
\begin{equation}
\label{eq:Maxwel} \phi_M(\vec{c}) = \pi^{-3/2}\exp(-c^2)\, ,
\end{equation}
and $S_k(c^2)$ are Sonine the polynomials. The coefficients $a_k^{(i)}$ have been computed in a few systems. In all examples they were rather small, e.g., in the case of dissipative collisions \cite{GarzohrenyaDuftyII} and in the case of reacting particles~\cite{Trizac_anih}. Therefore we shall use the Maxwell distribution function (\ref{eq:Maxwel}) in all further calculations. Integration of Eqs.~(\ref{eq:EB_wakes}) over $\vec{v}_k$ with the use of the Maxwell distribution function  (\ref{eq:Maxwel}) is rather straightforward, since all arising integrals are Gaussian. This integration, discussed in detail in the next section,  yields Eqs.~(1) and (2) of the main text.

\section{Derivation of the rate  coefficients }
To derive the rate equations (2) of the main text, we integrate the Boltzmann equation (21) of the main text over $\vec{v}_k$. Since $n_k=\int d \vec{v}_k f_k(\vec{v}_k ,t)$, the left hand side of the Boltzmann equation turns then into $dn_k/dt$ and gives the rate of change of the concentrations $n_k$. The right-hand side gives the contributions to $dn_k/dt$ from different parts of the collision integral. Since bouncing collisions and the heating term do not change the number of particles, we easily obtain (see also \cite{brillOUP}):
$$
\int d \vec{v}_k I^{\rm b}_k = \int d \vec{v}_k I^{\rm heat}_k=0.
$$
We use the Maxwellian distribution for the distribution function $f_k$,
$$
f_k(\vec{v}_k ,t)=\frac{n_k}{\pi^{3/2}v_{0,k}^3} e^{-v^2/v_{0,k}^2},
$$
where $v_{0,k}^2$ is the thermal velocity of aggregates comprised of $k$ monomers. Then the integral over $\vec{v}_k$ of  the second part of the aggregation integral, $I^{\rm agg,2 }_k$ (Eq.~(\ref{eq:Iagg})), may be written as
\begin{eqnarray}
\label{eq:Iagg2}
&&\int d \vec{v}_k I^{\rm agg,2 }_k =\sum_{j}\sigma_{kj}^2  \int d \vec{v}_k\int d \vec{v}_j \int d\vec{e} \, \Theta\left(-\vec{v}_{kj} \cdot \vec{e} \,
\right)     \nonumber \\
&&{} \qquad \qquad \times \left|\vec{v}_{kj} \cdot \vec{e} \,\right| f_k\left(\vec{v}_k \right)
f_j\left(\vec{v}_j \right) \Theta\left(E_{\rm agg}-E_{kj}\right)  \\
&&= \sum_{j}\frac{\sigma_{kj}^2n_kn_j}{\pi^3 v_{0,k}^3 v_{0,j}^3} \int d \vec{v}_k  d \vec{v}_j  d\vec{e} \, \Theta\left(-\vec{v}_{kj} \cdot \vec{e} \,
\right) \left|\vec{v}_{kj} \cdot \vec{e} \,\right|  \nonumber \\
&&{} \quad \qquad \times e^{-v_k^2/v_{0,k}^2 -v_j^2/v_{0,j}^2 }\Theta\left(E_{\rm agg}-\frac12\mu_{kj} v_{kj}^2 \right), \nonumber
\end{eqnarray}
where $\mu_{kj}=m_km_j/(m_k+m_j)$, $\vec{v}_{kj} =\vec{v}_k-\vec{v}_j$ and $E_{kj}=\mu_{kj} v_{kj}^2/2$. The integrals in the above equation are Gaussian and hence may be straightforwardly calculated. We perform this calculation for a particular pair $k$ and $j$. With the substitute
\begin{eqnarray}
  \vec{v}_k&=& \vec{u} +\vec{w} (\mu_{kj}/m_k -p_{kj}) \nonumber\\
  \vec{v}_j&=& \vec{u} -\vec{w} (\mu_{kj}/m_j +p_{kj})  \nonumber
\end{eqnarray}
where $p_{kj}= \mu_{kj} \left[ (m_k v_{0,k}^2)^{-1} - (m_j v_{0,j}^2)^{-1} \right] /
\left[ v_{0,k}^{-2} +v_{0,j}^{-2} \right]$, the above integral w.r.t. $ \vec{v}_k$, $ \vec{v}_j$ and $\vec{e}$ may be written as
$$
  \int d\vec{u}  d \vec{w}  d\vec{e}    \Theta\left(-\vec{w} \cdot \vec{e} \,
\right) \left|\vec{w} \cdot \vec{e} \,\right| e^{-au^2 -bw^2}
\Theta\left(E_{\rm agg}-\frac12\mu_{kj}w^2 \right), 
$$
where $a \equiv a_{kj}=v_{0,k}^{-2} +v_{0,j}^{-2}$ and $b \equiv b_{kj}=1/(v_{0,k}^{2} +v_{0,j}^{2})$ and we take into account that the Jacobian of transformation from $(\vec{v}_k, \, \vec{v}_j)$ to $(\vec{u}_k,\, \vec{w})$ is equal to unity.  Integration over $\vec{u}$ gives $(\pi/a)^{3/2}$. Integration over the unit vector $\vec{e}$ gives $4\pi$ and we are left with the integral over $\vec{w}$.
Integration over directions of the vector  $\vec{w}$ gives $\pi$, so finally we need to calculate the remaining integral:
$$
h_{kj}\!=\!\int_{0}^{\sqrt{\frac{2 E_{\rm agg}}{\mu_{kj}}}} \!\!\!\!dw w^3 e^{-bw^2}
\!=\!\frac{1}{2b^2}\left[1\!-\! e^{\frac{2bE_{\rm agg}}{\mu_{kj}}} \left( \!1+\! \frac{2bE_{\rm agg}}{\mu_{kj}} \right) \!\right].
$$
As the result we obtain:
\begin{equation}\label{eq:Iagg21}
 \int d \vec{v}_k I^{\rm agg,2}_k= \sum_{j}\frac{\sigma_{kj}^2n_kn_j}{\pi^3 v_{0,k}^3 v_{0,j}^3} \, 4\pi^2 \left( \frac{\pi}{a_{kj}} \right)^{3/2} h_{kj}.
\end{equation}
When we integrate  the first part of $I^{\rm agg,1}_k$ in Eq.~(\ref{eq:Iagg}) over $\vec{v}_k$, we observe that $\int d \vec{v}_k \delta(m_k \vec{v}_k-m_i \vec{v}_i-m_j \vec{v}_j ) =1$, since the other part of the integrand does not depend on $\vec{v}_k$. Than the remaining integration is exactly the same as for $I^{\rm agg,2}_k$, which have been already performed, therefore we find:
\begin{equation}\label{eq:Iagg_fin}
\int d \vec{v}_k I^{\rm agg,1}_k=   \frac12\sum_{i+j=k}\frac{\sigma_{ij}^2n_in_j}{\pi^3 v_{0,i}^3 v_{0,j}^3} \, 4\pi^2 \left( \frac{\pi}{a_{ij}} \right)^{3/2} h_{ij}.
\end{equation}
Turn now to the calculation of the integral over $\vec{v}_k$ of the fragmentation integral $I^{\rm frag}_k$. We notice that according to Eq.~(\ref{eq:qx}), the integration over $\vec{v}_k$ of $q_{ki} (\vec{v}_k, \vec{v}_i, \vec{v}_j)$ yields $x_k(i)$. Therefore for all three parts of $\int d \vec{v}_k I^{\rm frag}_k$ we need to compute the following integrals:

\begin{eqnarray}
\label{eq:Ifragvk1}
&&\int d\vec{v}_j \int \!d \vec{v}_i \int d \vec{e} \,
\Theta\left(-\vec{v}_{ij} \cdot \vec{e} \, \right)   \times  \\
   &&{}\qquad \qquad \times \left|\vec{v}_{ij} \cdot \vec{e} \,\right| f_i\left(\vec{v}_i\right)f_j\left(\vec{v}_j \right)
   \Theta\left(E_{ij}^{\rm n} - E_{\rm frag} \right), \nonumber
\end{eqnarray}
which have the same structure as the integrals in (\ref{eq:Iagg2}). The only difference is that instead of the factor $\Theta\left(E_{\rm agg}-E_{ij} \right)$ in (\ref{eq:Iagg2}) we have now $\Theta\left(E_{ij}^{\rm n} - E_{\rm frag} \right)$. Therefore calculations of the integrals (\ref{eq:Ifragvk1}) may be performed as previously. We apply the same transformation of variables as before and arrive at  the following integral,
$$
  \int d\vec{u}  d \vec{w}  d\vec{e}    \Theta\left(-\vec{w} \cdot \vec{e} \,
\right) \left|\vec{w} \cdot \vec{e} \,\right| e^{-au^2 -bw^2} \Theta\left(E_{\rm frag}
-\frac{\mu_{ij}w_n^2}{2} \right), 
$$
 where $w_n=(\vec{w} \cdot \vec{e})$, and the same notations as previously are used. Integration over $\vec{u}$ and  $\vec{e}$ gives again $4\pi(\pi/a)^{3/2}$. If we choose the direction of the vector $\vec{e}$ along $z$-axis (the direction of the $z$-axis is arbitrary), integration over $w_x$ and $w_y$ yields $(\pi/b)^{1/2} \cdot (\pi/b)^{1/2}$ and integration over $w_z$ leads to the integral:
$$
g_{ij}\!=\!\int_{-\infty}^{-\sqrt{ \frac{2 E_{\rm frag} }{\mu_{ij} }}}  dw_z |w_z| e^{-bw_z^2} \!=\!\frac{1}{2b}e^{-\frac{2 b E_{\rm frag}}{\mu_{ij}}}.
$$
Hence we obtain:
\begin{eqnarray}
\label{eqIfrag_vk}
&&\int d \vec{v}_k I^{\rm frag}_k = \\
 &&{}\quad = \frac12 \sum_{i,j\geq k+1} (x_{ki} + x_{kj})\frac{\sigma_{ij}^2n_in_j}{\pi^3 v_{0,i}^3 v_{0,j}^3} \, 4\pi \left( \frac{\pi}{a_{ij}} \right)^{3/2} \frac{\pi}{b_{ij}} g_{ij} \nonumber \\
 &&{} \qquad +\sum_{i=1}^k\sum_{j\geq k+1} x_{kj}\frac{\sigma_{ij}^2n_in_j}{\pi^3 v_{0,i}^3 v_{0,j}^3} \, 4\pi \left( \frac{\pi}{a_{ij}} \right)^{3/2} \frac{\pi}{b_{ij}} g_{ij}  \nonumber \\
   &&{} \qquad-\sum_{i }\left(1-  \delta_{k,1}\right) \frac{\sigma_{ki}^2n_kn_i}{\pi^3 v_{0,k}^3 v_{0,i}^3} \, 4\pi \left( \frac{\pi}{a_{ki}} \right)^{3/2} \frac{\pi}{b_{ki}} g_{ki}\nonumber
\end{eqnarray}
Combining Eqs.~(\ref{eq:Iagg21}), (\ref{eq:Iagg_fin}) and (\ref{eqIfrag_vk}),  and using the above definitions of the quantities $a_{ij}$, $b_{ij}$, $\mu_{ij}$, $h_{ij}$ and $g_{ij}$,  we arrive at the rate equations (2) with the according rate coefficients (1) of the main text. To write the rate coefficients $C_{ij}$ and $A_{ij}$ in the form of Eq.~(2) we take into account that the thermal velocity $v_{0i}$ is related to the mean square velocity $\left< {\bf v}_0^2 \right>$, termed here as the velocity dispersion, as
$$
\left< {\bf v}_0^2 \right>=\frac32 v_{0,i}^2,
$$
which is a direct consequence of the Maxwellian distribution.

\section{Self-gravity wakes and averaged kinetic equations}
\label{seq:wakes} Saturn's rings are not uniform but exhibit a large variety of structures \cite{colwell2009}.
One example are the self-gravity wakes \cite{salo1992,Colwell2006,Hedman2007,French2007}, arising from
self-gravitational instability, forming a transient and fluctuating pattern in the surface mass density of the
rings. These are canted relative to the azimuthal direction, with a typical length scale of about $L_w \sim
10^2$\,m, one Toomre critical wavelength~\cite{toomre1964}. Thus, to describe adequately particle kinetics
one needs, in principle, to take into account effects of dense packing and non-homogeneity.

Two important comments are to be done in this respect. First, due to the low velocity dispersion of particles
in the dense parts of the wakes, the collision duration is still significantly smaller than the time between
particle collisions. This implies the validity of the assumption of binary collisions, as the dominant
mechanism of particles' kinetics. Therefore a kinetic description in terms of the Enskog-Boltzmann equation
is possible~\cite{araki1986}.  Although this Markovian equation ignores memory effects in particle kinetics,
it may be still applicable, when the mean free path is comparable to, or even smaller then the particle
size~\cite{resibois1977}. Second, the characteristic length scale of  the density wakes, $L_w$, and the upper
cut-off radius (less then  $10 {\rm m}$) are well separated. This allows one to neglect variations of density and
distribution functions on the latter length scale and use a local approximation for the distribution function
of two particles at contact:
\begin{equation}
\label{eq:local}
 f_2(\vec{v}_k, \vec{r}-\vec{e}r_k, \vec{v}_l, \vec{r}+\vec{e}r_l,t) \simeq
g_2(\sigma_{lk}) f_k(\vec{v}_k, \vec{r},t)f_l(\vec{v}_l, \vec{r},t)
\end{equation}
Here $f_2$ is the two-particle distribution function corresponding to particles of radii $r_k$ and $r_l$, which have
a contact at point $\vec{r}$. The unit vector $\vec{e}$ joins the centers of particles and
$g_2(\sigma_{lk})$, with $\sigma_{lk} =r_l+r_k$, is the contact value of the pair distribution function. It may be well approximated by the corresponding equilibrium value for the hard sphere fluid; explicit expressions for
$g_2(\sigma_{lk})$ can be found, e.g., in~\cite{brillOUP}. In the local approximation, Eq.~(\ref{eq:local}),
the collision integrals depend only on local values at a particular space point $\vec{r}$. This significantly
simplifies the kinetic description of a high-density gas, since the density effects are taken into account in
this approach  by the multiplicative Enskog factor, $g_2(\sigma_{lk})$ only, leaving the structure of the
collision integrals unchanged. Therefore the Enskog-Boltzmann equation valid for the case of wakes reads:
\begin{eqnarray}
\nonumber
\frac{\partial}{\partial t} f_k\left(\vec{v}_k, \vec{r}, t \right) &+&\!\!\vec{v}_k\cdot
\vec{\nabla}  f_k\left(\vec{v}_k, \vec{r},  t \right) +\vec{F}_k(\vec{r}) \cdot
\frac{\partial}{\partial \vec{v}_k}  f_k\left(\vec{v}_k, \vec{r}, t \right) \nonumber \\
&=& I^{\rm agg}_{k}(\vec{r}) + I^{\rm b}_{k}(\vec{r})+I^{\rm frag}_{k}(\vec{r})
\label{eq:EB_wakes}
\end{eqnarray}
Here $ f_k\left(\vec{v}_k, \vec{r}, t \right)$ is the velocity distribution function of particles of size
$k$ which depends on the space coordinate $\vec{r}$. Further, $\vec{F}_k(\vec{r})$ is
the total gravitational force, acting on the particle of size $k$, which includes both the gravitational
force from the central planet as well as self-gravitation of the ring particles. In what follows we do not
need an explicit expression for this term. The collision integrals on the right-hand side of Eq.~(\ref{eq:EB_wakes})
have the same form as in the previous case of a uniform system, with the only difference, that they depend on
local parameters taken at a point $\vec{r}$, and that the collision cross-sections are re-normalized
according to the rule,
\begin{equation}
\label{eq:renorm} \sigma_{ij}^2 \, \longrightarrow  \, \sigma_{ij}^2 \, g_2(\sigma_{ij})\,
\end{equation}
which accounts for the high-density effects in the local approximation, see e.g.~\cite{brillOUP}. We do not need here the term $I^{\rm heat}_k(\vec{r})$ which mimics the heating for the model of a uniform gas, since Eq.~(\ref{eq:EB_wakes}) implicitly contains the spacial gradients and fluxes responsible for the heating.

Integrating the kinetic equation~(\ref{eq:EB_wakes}) over $\vec{v}_k$,  we find rate equations for the concentrations $n_k(\vec{r})$,
\begin{eqnarray}
\nonumber
&&\frac{\partial }{\partial t} n_k(\vec{r}) +  \vec{\nabla} \cdot
\vec{j}_k(\vec{r}) = \frac12 \sum_{i+j=k}C_{ij}(\vec{r}) n_i(\vec{r})n_{j}(\vec{r}) \\ \nonumber
&&-n_k(\vec{r})\sum_{i\geq1} C_{ki}(\vec{r}) n_i(\vec{r})-\sum_{i\geq1} A_{ki}(\vec{r})n_k(\vec{r})n_i(\vec{r})
\left(1- \delta_{k1} \right) \\ \nonumber
&&+\sum_{i=1}^k n_i (\vec{r}) \sum_{j\geq k+1}A_{ij}(\vec{r}) n_j(\vec{r}) x_k(j)+\\
&&+\frac12 \sum_{i,j\geq k+1}A_{ij}(\vec{r}) n_i (\vec{r})n_j(\vec{r}) \left( x_k(i)+x_k(j) \right),
\label{eq:Smoluchwakes}
\end{eqnarray}
where
$$
\vec{j}_k(\vec{r}) = \int \vec{v}_k \, f_k\left(\vec{v}_k, \vec{r}, t \right) d \vec{v}_k \, .
$$
is the macroscopic (hydrodynamic) flux associated with particles of size $k$. The important
feature of Eq.~(\ref{eq:Smoluchwakes}) is the spatial dependence of the  kinetic coefficients $A_{ij}$ and
$C_{ij}$. Although the structure of these coefficients coincides with that of kinetic coefficients in the
uniform system (apart from the trivial re-normalization, Eq.~(\ref{eq:renorm})), all quantities here are
local. Naturally,  the local velocity dispersion $\langle {\bf v}_i^2 \rangle (\vec{r})$ in the dense parts of the
wakes significantly differs from that of the dilute regions in between.

Now we average Eqs.~(\ref{eq:Smoluchwakes}) over a suitable control volume $V$, which contains a large number
of wakes. Applying then Green's theorem,
$$
\frac{1}{V} \int_V \vec{\nabla} \cdot \vec{j}_k  \, d \vec{r} = \frac{1}{V} \int_S \vec{j}_k \cdot d\vec{s} \sim \frac{S}{V}  \, ,
$$
we notice that the contribution of the term containing the flux $\vec{j}_k$ vanishes as $ S/V \to 0$ for large enough volume. As the result  we arrive at the set of equations with the
space-averaged quantities:
\begin{eqnarray}  \nonumber
&&\frac{d}{d t} \bar{n}_k =  \frac12\sum_{i+j=k}\bar{C}_{ij} \bar{n}_i\bar{n}_{j} -\bar{n}_k \sum_{i} \bar{C}_{ki}\bar{n}_i- \\ \nonumber
&&- \sum_{i} \bar{A}_{ki}\bar{n}_k\bar{n}_i \left(1- \delta_{k1} \right)
+\sum_{i=1}^k \bar{n}_i \sum_{j=k+1}^{\infty}\bar{A}_{ij} \bar{n}_j x_k(j)
\\
&&+ \frac12 \sum_{i,j\geq k+1} \bar{A}_{ij} \bar{n}_i \bar{n}_j \left( x_k(i)+x_k(j) \right)\,. \label{eq:SmoluchAver}
\end{eqnarray}
Here, by the definition,
$$
\bar{n}_k  =   \frac{1}{V} \int n_k(\vec{r}) d \vec{r}
$$
and
$$
\bar{C}_{ij} =  \frac{1}{ \bar{n}_i \bar{n}_j } \frac{1}{V} \int n_i(\vec{r}) n_j(\vec{r}) C_{ij}
(\vec{r}) d \vec{r} \, ,
$$
where $C_{ij} (\vec{r})$ are defined by Eqs.~(1) of the main text, with the local velocity dispersions $\langle {\bf v}^2_i \rangle
(\vec{r})$. Similar expression holds true for the coefficients  $ \bar{A}_{ij}$.

It is important to note that the coefficients $ \bar{C}_{ij}$ and $ \bar{A}_{ij}$ are density-weighted
quantities. Therefore the contribution to the average value is proportional to the local density. This
in turn implies that the values of these coefficients practically coincide with these for the dense
part of the wakes,
$$
\bar{C}_{ij} = {C}_{ij}^{\rm (dense \,\,part )} \qquad \qquad \qquad \bar{A}_{ij} = {A}_{ij}^{\rm (dense
\,\,part)}\, .
$$

Hence we conclude that the kinetic equations for the average concentrations of particles $\bar{n}_k$ coincide
with the previously derived equations for $n_k$ for the case of a uniform system. In what follows we will use
$n_k$,  $ C_{ij}$ and $A_{ij}$ for the notation brevity,  keeping although in mind that they correspond to
the average values, that are almost equal to these values in the dense part of the wakes.

\section{Estimates of the characteristic energies and the aggregates strength} Using the data for $\lambda$ reported in the main text, we perform here some estimates.
\\ \\
\noindent
\emph{Estimate of the fragmentation energy and of the aggregates strength.}
First we estimate the effective value of $ E_{\rm frag}$, assuming that $\left< B_{ij} E_{\rm agg} (R) \right> \gg 1$. According to Eqs.~(1) of the main text $C_{ij} \simeq \nu_{ij}$ and   $A_{ij} \nu_{ij} \exp(-B_{ij}E_{\rm frag})$ which yields
\begin{equation}\label{eq:lambAC}
  \lambda =A_{ij}/C_{ij} \simeq \exp(-B_{ij}E_{\rm frag}).
\end{equation}
We estimate the average value of $B_{ij}$. For two particles of equal mass $m_i=m_j$ with characteristic square velocity ${\bf v}_0^2$ this  quantity reads, according to Eqs.(1) of the main text:
\begin{equation}
\label{eq:Bij} B_{ij}= \frac32 \frac{(m_i^{-1}+m_j^{-1})}{ {\bf v}_0^2 } = \frac{9}{4 \pi \rho \phi {\bf v}_0^2} R^{-3}\, ,
\end{equation}
where $m_i=(4\pi/3) \rho \phi R^3$, $\rho=900 \, {\rm kg/m^3}$ is the material density of ice and $\phi=0.3$ is the approximate packing fraction of aggregates. Let us estimate the average value of $B_{ij}$ which we define as
$$
\left< B_{ij}(R) \right> = \frac{\int_{r_1}^{R_{c}} B_{ij}(R) F(R) dR}{\int_{r_1}^{R_c}  F(R) dR }\,.
$$
Here $F(R) \simeq {\rm const.} R^{-q}$ is the radii distribution function. It behaves as a  power-law with $q \simeq 3$ for $R < R_c$, with $R_c\gg r_1$ being the cutoff radius for the distribution. The averaging for $q=3$ yields,
$$
\left< B_{ij}\right> =  \frac{9}{10 \pi \rho \phi {\bf v}_0^2}r_1^{-3},
$$
and respectively  the average fragmentation energy,
\begin{equation}\label{eq:Egrag}
  \left< E_{\rm frag}\right> =-\log \lambda \left< B_{ij}\right>^{-1} =
  \frac53 \pi \rho \phi {\bf v}_0^2 r_1^3\log (R_c/r_1).
\end{equation}
Here we use Eq.~(20) of the main text, $R_c=r_1/\lambda^{2/3}$.
As it may be seen from the above equation, the estimate of the effective fragmentation energy sensitively depends on the monomer size $r_1$ and the characteristic square velocity ${\bf v}_0^2$. The plausible range for these values is $1 \leq r_1 \leq 10{\rm cm}$ and $0.01 \leq v_0 \leq 0.1 {\rm cm/s}$~\cite{cuzzi2009}.

To be consistent with the laboratory measurements we choose the particular values: $r_1=7{\rm cm}$ and $v_0=0.07{\rm cm/s}$ from the above intervals for $r_1$ and $v_0$ (other combinations of these parameters are also consistent with the laboratory data) to obtain $$\left< E_{\rm frag} \right>= 1.04 \cdot  10^{-6} {\rm J}.$$ This fragmentation energy is equal to the product of the fragmentation energy of a single contact between aggregates $E_b$ times the average number of contacts between monomers $\left< N_c \right>$ in the aggregates. It may be estimated as follows. If the radius of an aggregate, composed of monomers of radius $r_1$  is $R$ and the packing fraction is $\phi$,  the number of contacts reads,
$
N_c(R) = \left( R/r_1 \right)^3 \phi z_c ,
$
where $z_c$ is the average number of contacts with neighbours. For a random packing of spheres $z_c=4.7$~\cite{Peters2001} and  the averaged contact number is
$$
\left< N_c \right> = \frac{\int_{r_1}^{R_{c}} N_c(R) F(R) dR}{\int_{r_1}^{R_c}  F(R) dR }
=2\phi z_c \left( \frac{R_c}{r_1} \right)
$$
Here we again use $F(R) \simeq {\rm const.} R^{-q}$ with $q=3$. Hence we obtain the average contact energy for a monomer-monomer bond:
$$
E_b = \frac{\left< E_{\rm frag} \right> }{\left< N_c \right>}  =4.68\cdot 10^{-9} {\rm J}.
$$
Now we assume that the adhesive contacts between the monomers occurs in accordance with the {\it overlapping frost layer} model, as it follows from the laboratory measurements of~\cite{Hatzes1991,Hatzes1996}. This model of cohesion has been also used in the numerical simulation of the Saturn Rings, where aggregation and fragmentation processes have been taken into account~\cite{Perrine2011,Perrine2012}. The typical thickness of the frost layer is about $d=20 \mu$ which yields the estimate of the adhesion force $f_b$:
$$
f_b= \frac{E_b}{d} = 23.4\cdot 10^{-5}{\rm N} =23.4 {\rm dyne},
$$
in a good agreement with the laboratory data~\cite{Perrine2011,Perrine2012}, where the force of the order $30-50{\rm dyne}$ has been reported.

The contact area $S_b$ between the monomers, comprising an aggregate is equal to~\cite{Perrine2011}:
$$
S_b= \pi r_1^2 \beta(1-\beta/4),
$$
where the parameter $\beta=d/r_1$ has been introduced in~\cite{Perrine2011} and characterizes the ratio of the frost layer thickness and the particle radius. From the laboratory experiments follows that $\beta =10^{-3}$~\cite{Perrine2011}. This gives the estimate of the strength of  icy aggregates:
$$
P_b=\frac{f_b}{S_b} \sim 1.5 \cdot 10^6 {\rm Pa}.
$$
Indeed, this stress exists in the aggregates and keeps the constituents together. Therefore, it is reasonable to assume that any external stress smaller than $P_b$ would not destroy it. However, the applied external stress exceeding $P_b$, would most probably  cause  fragmentation. \\ \\
\emph{Estimate of the aggregation energy.}
Here we will estimate the value of $E_{\rm agg}$ for the overlapping frost layer model for particles contact. First we estimate the apparent adhesive coefficient $\gamma$, which is equal twice the is twice the surface free energy per unit area of a solid in vacuum. We use the laboratory data of~\cite{Hatzes1991,Hatzes1996} performed on the ice particles of radius $R_0=2.5{\rm cm}$. The characteristic force due to the frost layer of the thickness $d=10-30\mu$ was $f_0=30-50{\rm dynes}$, therefore the according energy, $f_0\cdot d$,  is about $40\cdot 10^{-5}{\rm N} \times  20 \cdot 10^{-6} {\rm m} = 8 \cdot 10^{-9} {\rm J}$. Therefore the apparent coefficient of adhesion reads,
$$
\gamma_{\rm eff} = \frac{f_0 d}{\pi R_0^2 \beta} = 0.0041{\rm J/m}.
$$
Now we would like to estimate the effective Young modulus and Poisson ratio of aggregates comprised of random packing of monomers, which have cohesive bonds as it follows from the  overlapping frost layer model.

Generally it may be shown that the effective Young modulus $Y_{\rm eff}$ and Poisson ratio $\nu_{\rm eff}$ of a random packing of spheres of radius $R$ that interact with a force $f(r)$ read:
$$
Y_{\rm eff} = \frac{3 \pi^2}{10} R^{-1/3} \phi z_c \left. \frac{df(r)}{dr}\right|_{r=r_{\rm eq}} \qquad \nu_{\rm eff} =\frac14,
$$
where $\phi$ is the packing fraction and $r_{\rm eq}$ is the equilibrium distance between the spheres' centres. Note that the effective Poison ratio does not depend on the microscopic detail and is determined  by the geometry of random packing only. To derive the above relations, one needs to consider two different types of deformation -- uniform compression (without shear) and uniform shear (without change of a volume). Then it is straightforward to compute the change of the system's elastic energy in terms of the deformations. This is to be done for the case of randomly packed spheres, with a given inter-particle force, and an average over configurations must be performed. The linear coefficients that relate the variation of the elastic energy to the respective deformation (for both types) yield the elastic coefficients, that is, the Young modulus and the Poisson ratio, as given in the above equation.

Using the experimental value of  $\frac{df(r)}{dr}=5.5\cdot 10^{4} {\rm dyne/cm}$ from~\cite{Hatzes1991} and $R=r_1 =7{\rm cm}$  we obtain the effective Young modulus, $Y_{\rm eff} = 557{\rm Pa}$ for the aggregate material, treated as a elastic continuum medium.

With the obtained effective values for the Young modulus, Poisson ratio and adhesive coefficient we can apply the effective JKR model, treating the colliding aggregates as  continuum bodies with the effective material parameters (that is, ignoring the discrete structure of the aggregates).  Then the threshold energy of aggregation for two particles of radii $R_i$ and $R_j$ has the following form \cite{bri2}:
\begin{equation}
\label{eq:Eagg} E_{\rm agg} = q_0 \left(\pi^5 \gamma^5 R_{\rm eff}^4 D^2\right)^{1/3} \, .
\end{equation}
Here $q_0=1.457$ is a constant, $\gamma$ is the adhesion coefficient, $D=(3/2)(1-\nu^2)/Y$, where $Y$ and $\nu$ are respectively the Young modulus and Poisson ratio and $R_{\rm eff} =R_iR_j/(R_i+R_j)$. In the case of interest we need to use the effective values for all the parameters. To estimate $\left<E_{\rm agg}\right>$ we apply the above expression for particles of equal radius to obtain,
$$
\left<E_{\rm agg}\right> = \frac{ \int_{r_1}^{R_c} E_{\rm agg}(R)F(R)dR}{\int_{r_1}^{R_c} F(R)dR} = 3 q_0 \left(\pi^5 \gamma_{\rm eff}^5 (r_1/2)^2 D_{\rm eff}^2 \right)^{1/3}
$$
where $D_{\rm eff} = (3/2)(1-\nu_{\rm eff}^2)/Y_{\rm eff}$. This gives $$\left<E_{\rm agg}\right> =6.21\cdot 10^{-7} {\rm J},$$ that is,
$$\left<E_{\rm frag }\right> =1.68\left<E_{\rm agg}\right>,$$
which implies that the aggregation and fragmentation energies are rather close.

Finally we compute $\left<B_{ij}E_{\rm agg}(R)\right>$. Since we have already computed $\left<B_{ij}(R)\right>$ and $\left<E_{\rm agg}(R)\right>$, we just apply the approximation:

$$
\left<B_{ij}E_{\rm agg}(R)\right> \approx \left<B_{ij}(R)\right>  \left<E_{\rm agg}(R)\right>,
$$
which yields $\exp\left(-\left<B_{ij}E_{\rm agg}(R)\right>\right)\approx 0.02$ and justifies the approximation $A_{ij}/C_{ij} = \exp(-B_{ij} E_{\rm frag})$ used in the main text.

\section{Universality of particles size distribution for steep distribution of debris}
As it has been already mentioned in the main text, distribution of debris in a collision obeys in its main
part a power-law. That is, if an aggregate of size $i$ suffers a disruption in an impact a plenty of fragments
of size $k < i$ appear. Let $x_k(i)$ denote the number of fragments of size $k$; the power-law fragment
distribution implies that $x_k(i) \sim k^{-\alpha}$ in the main part of the distribution. This allows to
quantify the pre-factor of the distribution,  $x_k(i) = B(i) k^{-\alpha}$ from the normalization condition,
that is, from the condition that the total mass of all debris is equal to the mass of the parent body.
Although we have a discrete mass spectrum of debris, $m_k =m_1k$, $k=1,2, \ldots$, for $i \gg 1$ one can
approximate summation by integration to obtain,
\begin{equation}
\label{eq:norm} i \simeq   \int_{1}^{i-1} B k^{-\alpha} k dk =  \left\{
    \begin{array}{ll}
         \beta\frac{ B}{(2-\alpha)} \left[(i-1)^{2-\alpha} -1 \right]   & \mbox{if } \alpha \neq 2 \\
        \beta_1 B \log (i-1)  & \mbox{if } \alpha =2,
    \end{array}
\right.
\end{equation}
where the factors $\beta$ and  $\beta_1$ stand for an approximate correction when the summation is
approximated  by the integration (see next section for more detail). This yields for $i \gg 1$
\begin{equation}
B(i) \sim   \left\{
    \begin{array}{ll}
          i^{\alpha-1}   & \mbox{if  } \alpha < 2 \\
         i (\log i )^{-1}   & \mbox{if  } \alpha =2 \\
         i & \mbox{if  } \alpha >2 \,.
    \end{array}
\right.
\end{equation}
Now we perform analysis of the general system of equations (2) in the main text to show that under certain
conditions the solution to Eqs. (2) (fragmentation with a particular debris-size distribution) coincides with
the solution to Eqs. (5)--(6) (complete fragmentation into monomers). First, we notice that if $x_k(i) \sim i/
k^{\alpha}$, which holds true for $\alpha >2$,  the equations  for monomers are identical (up to a factor at
the coefficients $A_{ij}$) for the both models. Next, we write Eqs. (2) as
\begin{eqnarray}
\label{kinpow} \frac{dn_k}{dt}= K_1-K_2-K_3+K_4+K_5 \,,
\end{eqnarray}
where
\begin{eqnarray}
\nonumber &&K_1=\frac12 \sum_{i+j=k} C_{ij}n_in_j\\\nonumber && K_2=\sum_{i=1}^{\infty}C_{ik}n_in_k\\\nonumber
&&K_3=\sum_{i=1}^{\infty}A_{ki}n_in_k \left(1-\delta_{k1}\right)\\\nonumber &&K_4=\sum_{i=1}^k
\sum_{j=k+1}^{\infty} A_{ij}n_in_j x_k(j)\\\nonumber &&K_5=\frac{1}{2} \sum_{i,j \geq k+1}  A_{ij}n_in_j
\left[x_k(i)+x_k(j) \right] \,. \nonumber
\end{eqnarray}
In these notations Eqs. (5) for the case of decomposition into monomers read
\begin{eqnarray}
\label{kinpow1} \frac{dn_k}{dt}= K_1-K_2-K_3 \,,
\end{eqnarray}
that is, the two models differ by the two terms $K_4$ and $K_5$ only. Now we estimate the relative magnitude
of the terms $(K_1-K_2-K_3)$ and $K_4$ and $K_5$ using the scaling approach. We assume that under certain
conditions, the distribution of aggregate concentrations in a steady state (when $\dot{n}_k =0$) has the form,
$$
n_k  \sim k^{-\gamma} e^{-ak}\,,
$$
the same as for the case of decomposition into monomers. We perform the analysis for the generalized product kernels
$C_{ij}=(ij)^{\mu}$ and $A_{ij}=\lambda C_{ij}$. We are interested in the scaling regime, $k \gg 1$, and we focus on the power-law domain where $ka \ll 1$. Additionally,  we assume that $1 < \gamma -\mu <2 $; we will check all the assumptions a posteriori.  Approximating again the summation by the integration we obtain for the first term $K_1$:
\begin{eqnarray}
\nonumber &&K_1 \sim \sum_{i+j=k} (ij)^{\mu -\gamma} e^{-ak}\sim \int_1^{k-1} i^{\mu-\gamma}
(k-i)^{\mu-\gamma}di  \\\nonumber
&&\sim    k^{\mu-\gamma} \int_1^{k/2} i^{\mu-\gamma}\left( 1-\frac{i}{k} \right)^{\mu-\gamma} di \\
&&\sim k^{\mu-\gamma} \left[ i^{\mu-\gamma+1} - b \, \frac{i^{\mu-\gamma+2}}{k} \ldots \right]_1^{k/2}
\sim k^{\mu -\gamma} ,
\label{eq:1st}
\end{eqnarray}
where $b=(\mu-\gamma)(\mu-\gamma+1)(\mu-\gamma+2)^{-1}$. Here  we take into account the symmetry of the
integrand around $k/2$, make an expansion of the factor $(1-i/k)^{\mu-\gamma}$ and keep only the leading term
in the obtained series. Now we evaluate the second and the third terms:
\begin{eqnarray}
\label{eq:2nd} K_2 + K_3 \sim (1+\lambda) k^{\mu -\gamma} e^{-ak} \int_1^{\infty} i^{\mu -\gamma} e^{-ai} di
\sim  k^{\mu -\gamma}.
\end{eqnarray}
Similarly, we find for the fourth term:
\begin{eqnarray}
&&K_4 \sim \lambda \sum_{i=1}^k i^{-\gamma} e^{-ai} \sum_{j=k+1}^{\infty}(ij)^{\mu} j^{-\gamma} e^{-aj}B(j) k^{-\alpha}\nonumber\\
&&  \sim k^{-\alpha} \int_{1}^{k} i^{\mu-\gamma } e^{-ai} di \int_k^{\infty} j^{\mu-\gamma } e^{-aj} B(j) dj.
\nonumber
\end{eqnarray}
Using  $B(j)$ from Eq.~(\ref{eq:norm}), it is straightforward to show, that the forth term scales as
\begin{equation}
\label{eq:4thscal} K_4  \sim \left\{
    \begin{array}{ll}
         k^{-\alpha} & \mbox{if } \, \, \alpha > \gamma -\mu \,\,,\, \alpha\neq 2   \\
         k^{-\alpha}\left( \log k \right)^{-1} & \mbox{if } \, \, \alpha =2  \\
         k^{-\alpha}\left| \log ka \right| \!\!& \mbox{ if} \, \, \alpha = \gamma -\mu \\
         k^{\mu-\gamma} & \mbox{if } \, \, \alpha < \gamma -\mu
    \end{array}
\right.
\end{equation}
Analogously,  the fifth term is estimated to give
\begin{eqnarray}
&&K_5\sim  \sum_{i,j \geq k} (ij)^{\mu} (ij)^{-\gamma} e^{-a(i+j)} \left[ B(i)+B(j) \right]k^{-\alpha} \nonumber\\
&& \sim   k^{-\alpha} \int_{k}^{\infty } di i^{\mu -\gamma} e^{-ai} \int_{k}^{\infty } j^{\mu -\gamma}
e^{-aj} B(j) dj. \nonumber
\end{eqnarray}
Finally, we get:
\begin{equation}
\label{eq:5thscal} K_5  \sim \left\{
    \begin{array}{ll}
         k^{\mu+1-\alpha-\gamma} & \mbox{if } \, \, \alpha > \gamma -\mu\,\,,\, \alpha\neq 2  \\
         k^{\mu+1-\alpha-\gamma}\left( \log k \right)^{-1} & \mbox{if } \, \, \alpha =2  \\
         k^{\mu+1-\alpha-\gamma}\left| \log ka \right| \!\!& \mbox{ if} \, \, \alpha = \gamma -\mu \\
         k^{1-2(\gamma -\mu)} & \mbox{if } \, \, \alpha < \gamma -\mu
    \end{array}
\right.
\end{equation}
Comparing Eqs.~(\ref{eq:2nd}) and  (\ref{eq:1st}) with  Eqs.~(\ref{eq:4thscal}) and (\ref{eq:5thscal}) we
conclude that the forth and fifth terms of the basic Eq.~(2) are negligibly small for $k \gg 1$ as compared
with the first, second and third terms of this equation, provided $\alpha > (\gamma -\mu)$ under the condition
$1<\gamma-\mu <2$. If we additionally take into account that for $\alpha >2$ the equations for the monomer
concentration coincide for the two models, we conclude that if $\alpha > {\rm max} \left\{ \gamma -\mu, 2
\right\}$, the steady-state size distribution of aggregates for the case of complete decomposition into
monomers and for the power-law decomposition would be the same in the domain $k \gg 1$ and $ka \ll 1$. Since
for the case of monomer decomposition $\gamma = 3/2 +\mu$ and $a=\lambda^2$, the condition $1< \gamma-\mu < 2$ holds true, and $ka \ll 1$ is fulfilled even for large $k$ if $\lambda \ll 1$. Hence it is expected that for a
steep size distribution of debris with $\alpha > \alpha_0 =2$ the steady-state distribution
\begin{equation}
\label{eq:n_k_univ} n_k \sim k^{3/2+\mu}
\end{equation}
is \emph{universal}, that is, it does not depend on the particular value of $\alpha$.

Moreover, the same conclusion of the universality of the distribution (\ref{eq:n_k_univ}) holds true for any
functional form of a steep distribution of debris size. If we write it as $x_k(i) = B(i)\phi(k)$, where the
function $\phi(k)$ is steep enough, so that
$$
\int_1^{i-1} \phi(k)k dk \simeq \int_1^{\infty} \phi(k)k dk
=C^{-1},
$$
the pre-factor $B(i)$ reads, $B(i)=Ci$. Then for any function $\phi (k)$ satisfying the condition, $\phi(k)
\ll k^{-3/2}$ for $k \gg 1$ the resultant distribution of aggregates will have the universal form
(\ref{eq:n_k_univ}). This  has been confirmed numerically for the exponential distribution of debris, see Fig.~2 in the main text.

\section{Particle size distribution in the presence of collisions with erosion}
Here we consider a more general case when in addition to the disruptive collisions, that completely destroy aggregates, there exist collisions with an erosion. In the erosive collisions a small fraction of the colliding particle mass is chipped off~\cite{Guettler2010,BlumErosion2011}.  To understand, what is the impact of the  erosive collisions on the particle size distribution, we consider a simplified model: A disruptive collisions occurs, if the kinetic energy of the relative normal motion  $E_{ij}^{\rm n}$ exceeds $E_{\rm frag}$, while the erosive collision takes place if $E_{ij}^{\rm n}$ exceeds smaller energy, $E_{\rm eros}$. That is, the condition for the erosive collision reads, $E_{\rm eros} \leq E_{ij}^{\rm n} < E_{\rm frag}$. We assume that in an erosive collision a piece of a \emph{fixed} size is chipped off from one of the colliding partners. Let the chipped-off piece always contains $l$ monomers. This piece may be further decomposed into smaller fragments. Below we consider two limiting cases, when the chipped-off piece remains intact and when it breaks into $l$ monomers. Performing the same derivation steps as for the fragmentation without erosion we arrive at the rate equations, that may be written  for the both cases uniformly. For simplicity we consider here the case of complete decomposition into monomers. Then the equation for monomers reads,
\begin{eqnarray}
\label{eq:n1}
  \frac{dn_1}{dt} &=& -n_1\sum_{i\ge 1} C_{1i}n_i +\frac{\lambda}{2}\sum_{i,j \geq 2} (i+j) C_{ij}n_in_j   \\
   &+&  \lambda n_1 \sum_{i \geq 2 } C_{1i}i n_i + \epsilon \lambda_e l \sum_{i \geq 1} \sum_{j \geq l+2}  C_{ij}n_in_j, \nonumber
\end{eqnarray}
where $\epsilon =1$ if the chipped-off piece disintegrates into monomers and $\epsilon =0$ if it remains intact. For $  k \geq l+2$  we obtain,
\begin{eqnarray}
\label{eq:nkl2}
  \frac{dn_k}{dt} &=& \frac{1}{2}\sum_{i+j =k} C_{ij}n_in_j  -(1+\lambda)  \sum_{i \geq 1 } C_{ik}n_i n_k \\
   &+&  \lambda_e \sum_{i \geq 1 } C_{i\, k+l}n_i n_{k+l} -\lambda_e \sum_{i \geq 1 } C_{i k}n_i n_{k},  \nonumber
   \end{eqnarray}
and, correspondingly  for $ l+1 \geq k \geq 2 $:
\begin{eqnarray}
\label{eq:nkl1}
  \frac{dn_k}{dt} &=& \frac{1}{2}\sum_{i+j =k} \!C_{ij}n_in_j  -(1+\lambda)  \sum_{i \geq 1 } C_{ik}n_i n_k \\
   &+&  \lambda_e \sum_{i \geq 1 } \!C_{i\, k+l}n_i n_{k+l} \!+ \!(1\!-\!\epsilon) \delta_{k,l} \lambda_e \!\! \sum_{i \geq 1} \!\sum_{j \geq l+2 } C_{ij}n_i n_{j}. \nonumber
\end{eqnarray}
The new coefficient $\lambda_e$ characterizes the relative frequency of the erosive and aggregative collisions:
\begin{equation}
\label{eq:lambda1}
  \lambda_e= \frac{e^{-B_{ij} E_{\rm eros}}\left(1-e^{-B_{ij} (E_{\rm frag} -E_{\rm eros})} \right)}{1-(1+B_{ij}E_{\rm agg})e^{-B_{ij}E_{\rm agg}}},
\end{equation}
where the coefficients $B_{ij}$ are defined in Eq.~(1) of the main text.\footnote{Note that in the above model of an erosive collision with the chip-off of an intact piece, the monomers do not appear. They are produced in the disruptive collisions.} We illustrate the derivation of size distribution for the case of complete disintegration of the chipped-off  piece, which corresponds to $\epsilon=1$ in the above equations; the case of $\epsilon =0$ is more simple and yields qualitatively same results. Similar to the analysis of the main text, we assume that the conditions that keep $\lambda$ and $\lambda_e$ constant are fulfilled. Moreover, we also assume that these constants are small and are of the same order of magnitude, that is, $\lambda_e =\alpha \lambda$, where $\alpha$ is of the order of unity, $\alpha \sim 1$. Since in erosive collisions the value of $l$ is not very large, and $\lambda_e$ is small, we further assume that $l \lambda_e \ll 1$.

For notation simplicity we perform the analysis for the generic case of constant rate coefficients, $C_{ij}=1$. Similar to the main text, the analysis may be undertaken for the more general case of $C_{ij}=(ij)^{\mu}$. We are looking for the steady-state distribution, when $dn_k/dt=0$. With $C_{ij}=1$ and $\epsilon=1$ the above equations read for a steady-state:
\begin{eqnarray}
\label{eq:n1Cij1}
  0 &=& -n_1\sum_{i\ge 1} n_i +\frac{\lambda}{2}\sum_{i,j \geq 2} (i+j) n_in_j   \\
   &+&  \lambda n_1 \sum_{i \geq 2 } i n_i +  \lambda_e l \sum_{i \geq 1} \sum_{j \geq l+2}  n_in_j, \nonumber \\
\label{eq:nkl1Cij1}
  0 &=& \frac{1}{2}\sum_{i+j =k} n_in_j  -(1+\lambda)  \sum_{i \geq 1 } n_i n_k \\
  &+&  \lambda_e \sum_{i \geq 1 } n_i n_{k+l} -  \theta(k\!-\!l-\!2) \lambda_e \sum_{i \geq 1} n_i n_{k}, \nonumber
\end{eqnarray}
where the Heaviside step function $\theta(k)=1$ for $k\ge 0$ and $\theta(k)=0$ for $k < 0$.

Now we apply the generation function technique, with a slightly different definition of this function, ${\cal N}(z) =\sum_{k\geq 1} n_k z^{k+l}$, so that the concentrations $n_k$ are the coefficients of ${\cal N}(z)$ at $z^{k+l}$. It is straightforward to show, that the generation function satisfies the quadratic equation
$$
{\cal N}(z)^2 - 2 A(z){\cal N}(z) +B(z)=0
$$
with
\begin{eqnarray}
\label{eq:Az}
  A(z) \!\!&=& \!\!(1+(1+\alpha) \lambda) Nz^l- \alpha \lambda N \\
  \label{eq:Bz}
  B(z) \!\! &=& \!\! 2N\left( (1+ \lambda)n_1z^{2l+1} - \alpha \lambda (1-z^l) G(z)\right),
\end{eqnarray}
where $G(z)=\sum_{i=1}^{l+1} n_i z^i$.

Since $\lambda$ is small, we analyze the expansion of ${\cal N}(z)$ in terms of $\lambda$. We use this expansion for the total number of aggregates, $N=\sum_{i \geq 1} n_i$, and for the concentrations:
\begin{eqnarray}
  N &=& N^{(0)}+\lambda N^{(1)} +\lambda^2 N^{(2)}+\lambda^3 N^{(3)} +\ldots \\
  n_i &=& n_i^{(0)}+\lambda n_i^{(1)} +\lambda^2 n_i^{(2)}+\lambda^3 n_i^{(3)} +\ldots
\end{eqnarray}
Substituting the above expansion for $n_i$ into Eqs.~(\ref{eq:n1Cij1}) and (\ref{eq:nkl1Cij1}) and summing these equations up, we find $N^{(0)}=n_1^{(0)}=0$ and
\begin{eqnarray}
\label{eq:Nexp}
   N^{(1)} \!&=&\! 2, \quad N^{(2)}\! =\!-4+2 \alpha K \quad N^{(3)} \!=\!8-4\alpha K \\
 \label{eq:niexp}
   n_1^{(1)} \!&=&\! 1, \!\! \!\! \qquad n_1^{(2)}\! =\!-1+\alpha K \qquad n_1^{(3)} \!=\!1-\alpha K
\end{eqnarray}
with $K=l\sum_{j \geq l+2} n_j^{(1)}$.

Now we can analyze
\begin{equation}\label{eq:calNzsqrt}
{\cal N}(z) = A(z) - \sqrt{ A^2(z) -B(z)}
\end{equation}
in different orders of $\lambda$. Keeping only terms of the order of $\lambda$ we find:
$$
{\cal N}(z) =\lambda z^l \left( N^{(1)} -  N^{(1)}\sqrt{1-2zn_1^{(1)}/N^{(1)}}\right).
$$
With $n_1^{(1)}=1$ and $N^{(1)}=2$ we obtain,

\begin{equation}\label{eq:NGamma}
  {\cal N}(z)= \lambda \sum_{k=1}^{\infty}\frac{\Gamma(k-1/2)}{\sqrt{\pi} \Gamma(k+1)} z^{k+l},
\end{equation}
which implies,  that in the first order in $\lambda$ the concentrations of the aggregates $n_k $ behave for $k \gg 1$ as
$$
n_k \simeq \frac{\lambda}{\sqrt{\pi}} k^{-3/2},
$$
that is, we obtain exactly the same power-law behavior as previously,  for the case of solely disruptive collision without erosion. To find the exponential cutoff for this power-law dependence, one need to consider next-order terms in $\lambda$.

To do this, we will focus on the part of ${\cal N}(z)$ given by Eq.~(\ref{eq:calNzsqrt}), corresponding to the terms $\sim z^m$ with $m \gg l,1$. Since $A(z)$ contains only terms up to  $z^{l+1}$, we analyze the behavior of $\sqrt{ A^2(z) -B(z)}= \sqrt{f(z)}$.

Obviously, $f(z)$ is a polynomial of the degree $2l+1$ and may be written as $f(z)=(z_1-z)(z_2-z) \ldots (z_{2l+1}-z)$, where $z_1 \leq z_2 \leq \ldots z_{2l+1}$ are the roots of $f(z)$. We are looking for the expansion of $\sqrt{f(z)}$ in term of $z$ in the vicinity of $z=0$. Moreover, we are interested in the expansion coefficients for $z^m$ with $m \gg 1$. For $m \gg 1$ the expansion coefficients at $z^m$ of $\sqrt{f(z)}$ coincide with the expansion coefficients of the function, $\sqrt{|z_1 f^{\prime}(z_1)|} \, \sqrt{1-z/z_1}$, which may be written as (see e.g.~\cite{AnalyticCominatoric}):
\begin{equation}\label{eq:sqrtfz}
  \sqrt{|z_1 f^{\prime}(z_1)|} \, \sqrt{1-\frac{z}{z_1}}  =
     \sqrt{\frac{|z_1 f^{\prime}(z_1)|}{4\pi}}\, \sum_{k=1}^{\infty} \frac{\Gamma\left(k-\frac12 \right)}{ \Gamma(k+1)} \left(\frac{z}{z_1} \right)^{k}.
\end{equation}
As it follows from Eq.~(\ref{eq:NGamma}), the closest to $z=0$ root of $f(z)$ is equal to
one ($z_1=1$) in the first order in $\lambda$. Therefore we need to find the next order corrections to $z_1$ in powers of $\lambda$:
$$
z_1=1 + \lambda \omega_1 + \lambda^2 \omega_2.
$$
Substituting into $f(z) = A^2(z) -B(z)=0$ the expansions  for $N$ and $n_i$, given by  Eqs.~(\ref{eq:Nexp}), (\ref{eq:niexp}) and the above expansion for $z_1$, one can find the coefficients $\omega_1$ and $\omega_2$:
$$
\omega_1=0, \qquad \qquad \omega_2=1,
$$
which means that $z_1 = 1 + \lambda^2 + \ldots$. Therefore, the  high-order terms of ${\cal N} (z)$, that contain  $z^k$ with $k \gg l,1$, behave as:
$$
\frac{\Gamma(k-\frac12)}{\Gamma(k+1)} \left(\frac{z}{z_1} \right)^{k}
\sim k^{-3/2} e^{-k\log z_1} \sim k^{-3/2}e^{-\lambda^2k}.
$$
Such dependence of the concentrations $n_k$ on $k$  coincides with the one for the model of disruptive collisions discussed  in the main text. This is confirmed by the numerical solution of the respective rate equations, see Fig.~1.

The case of $C_{ij}=(ij)^{\mu}$ may be analysed analogously; in this case one obtains the exponent $-(\mu +3/2)$ instead of $-3/2$, that is, again the same behavior as for purely disruptive collisions. Hence we conclude that the presence of the collisions with an erosion does not change qualitatively the size distribution of particles in planetary rings.

\section{Efficient numerical solution of the kinetic equations}

We use the Euler's method for the numerical solution of the kinetic equations.  This method is rather suitable
for the case of interest, since we search stationary, continuous and smooth solutions. The first problem in
the numerical analysis of the infinite number of rate equations is the conservation of mass. Indeed, in any
real simulation one can handle only a \emph{finite} number of equations say $N$, which describe evolution of
particles of size $1,\, 2,\, \ldots N$ (a particle of size $k$ has mass $m_k=m_1 k$). These equations have
both aggregation and fragmentation terms. In particular, they have a term which describes aggregation of
particles of size $i <N$ and $j <N$, resulting in an aggregate of size $i+j >N$. Since the system of $N$
equations does not account for particles larger that $N$, such processes would effectively lead to the leak of
particles' mass and violation of the mass conservation. To preserve the mass conservation we assume that all
collisions of particles of mass $i$ and $j$ are \emph{fragmentative} if $i+j \geq N$. We have checked that
this assumption does not lead to any noticeable distortion of the numerical solution of the rate equations
$n_k$,  if $k$ is smaller than some fraction of $N$.

Another problem is  to handle efficiently  a large number of equation, say up to $\sim 10^6$. One possible
way is an application of the coarse-graining, that is, grouping concentrations $n_k$ -- $n_{k+l}$ into
coarse-grained variables $\tilde{n}_K$ with increasing $l$ as $k$ grows. In the case of interest, however, we
have a drastic variation of the functional dependence of $n_k(k)$, which changes from a power-law to the
exponential decay. This hinders an effective application of the coarse-graining and we need to keep
explicitly all individual concentrations. Hence we have to work with a large number of equations, which is
computationally costly. To speed up the computations we  have developed a  recursive procedure.

In the case of fragmentation into monomers the system of kinetic equations has the following form:
\begin{eqnarray}
\nonumber &&\frac{d n_1}{dt} = - n_1\sum_{j=1}^{N} n_j +\lambda(1-n_1)\sum_{j=1}^{N} n_j\\ &&...\\ \nonumber
&&\frac{dn_{k}}{dt}=\frac{1}{2}\sum_{i+j=k}C_{i,j}n_{i}n_{j}-(1+\lambda)\sum_{i=1}^{N}C_{i,k}n_{i}n_{k}\\
\nonumber &&\frac{dn_{k+1}}{dt}=\frac{1}{2}\sum_{i+j=k+1}C_{i,j}n_{i}n_{j}-(1+\lambda)
\sum_{i=1}^{N}C_{i,k+1}n_{i}n_{k+1} \\
&&...  \nonumber \label{eq:sys}
\end{eqnarray}
Taking into account, that we search for a stationary solution, $dn_{k+1}/dt=0$, we  obtain  for the number
density $n_{k+1}$:
\begin{equation}
\label{eq:begin} \frac{1}{2(1+\lambda)}\sum_{i+j=k+1}C_{i,j}n_{i}n_{j}
-\sum_{i=1}^{N}C_{i,k+1}n_{i}n_{k+1}=0\,.
\end{equation}
The first sum in Eq.~(\ref{eq:begin}) contains only $n_i$ with $i\leq k$, while we write the second sum as
\begin{eqnarray}\nonumber
\sum_{i=1}^{N}C_{i,k+1}n_{i}n_{k+1} &=&n_{k+1}\sum_{i=1}^{k}C_{i,k+1}n_{i}+C_{k+1,k+1}n_{k+1}^2\\
&+&n_{k+1}\sum_{i=k+1}^{N}C_{i,k+1}n_{i} \,. \label{eq:javn}
\end{eqnarray}
Now we use the properties of the kinetic kernel $C_{ij}$ and the steady-state distribution $n_k=n_k(m_k)$,
which we assume to be decreasing function of $k$.  Namely, we assume that the coefficients $C_{ij}$ increase
with $i$ and $j$ at a smaller rate than the rate at which $n_k$ decreases with $k$. That is, we assume that
for $k \gg 1$ the following condition holds true:
\begin{equation}\label{eq:cond}
\sum_{i=1}^{k}C_{i,k+1}n_{i}>>\sum_{i=k+1}^{N}C_{i,k+1}n_{i} \,.
\end{equation}
This allows to neglect the last sum in Eq.~(\ref{eq:javn}) and obtain the quadratic equation for $n_{k+1}$:
\begin{equation}
\label{eq:sqeq} C_{k+1,k+1}n_{k+1}^{2} +n_{k+1}\sum_{i=1}^{k}C_{i,k+1}n_{i}  - \sum_{i+j=k+1}
\frac{C_{i,j}n_{i}n_{j}}{2(1+\lambda)}=0 \,.\nonumber
\end{equation}
Solving the above  equation and choosing the positive root, we arrive at the recurrent relation for the
concentrations $n_k$:
\begin{eqnarray}
\label{eq:req} &&n_{k+1} = \frac{ \sqrt{D}- \sum\limits_{i=1}^{k}C_{i,k+1}n_{i}  }{2C_{k+1,k+1}} \\
&&D= \frac{2C_{k+1,k+1}}{(1+\lambda)}\sum\limits_{i+j=k+1}C_{i,j}n_{i}n_{j}+
\left(\sum\limits_{i=1}^{k}C_{i,k+1}n_{i}\right)^2\,. \nonumber
\end{eqnarray}
Using the recurrent relation (\ref{eq:req}) one can significantly accelerate computations. This may be done
as follows:~First, one solves explicitly the system of rate equations (\ref{eq:sys}) for $k \ll N$, choosing
the value of $k$ to fulfil the condition (\ref{eq:cond}). Then the concentrations $n_i$ with $k < i \leq N$
may be straightforwardly obtained from the recurrence (\ref{eq:req}). Performing numerical solution of the
rate equations with different kernels directly, and with the use of the recurrence (\ref{eq:req}), we proved
the efficiency and accuracy of the above accelerating approach.

\section{Numerical calculation of the distribution of fragments}

In the numerical solution we calculate $x_k\left(i\right)$ using the mass conservation  and taking into
account the discreteness of the distribution of particles:

\begin{equation}
kx_k\left(i\right) = B\left(i\right) \int_{k-1/2}^{k+1/2} k_1k_1^{-\alpha}dk_1
\end{equation}

Computing the integral, we find:\\

if $\alpha \neq 2$:
\begin{equation}
x_k\left(i\right) = \frac{B\left(i\right)}{k}\frac{1}{2-\alpha} \left[\left(k+1/2\right)^{2-\alpha}-\left(k-1/2\right)^{2-\alpha}\right]\;
\end{equation}

if $\alpha = 2$:
\begin{equation}
x_k\left(i\right) = \frac{B\left(i\right)}{k}\left[\ln\left(k+1/2\right)-\ln\left(k-1/2\right)\right]\,.
\end{equation}

Note, that $x_k\left(i\right) \rightarrow B\left(i\right) k^{-\alpha}$ for $k >> 1$, when the discreteness of the system becomes insignificant.

Here $B\left(i\right)$ represents a normalization constant, which can be computed from:

\begin{equation}
B\left(i\right) \int_{1/2}^{i-1/2}kk^{-\alpha}dk = i\,.
\end{equation}

Thus we obtain:\\

if $\alpha \neq 2$:
\begin{equation}
B\left(i\right) = \frac{i\left(2-\alpha\right)}{\left(i-1/2\right)^{2-\alpha}-\left(1/2\right)^{2-\alpha}}\,,
\end{equation}
so that $B(i) = 2^{2-\alpha} (\alpha-2) i$ for $i \gg 1$;

 if $\alpha = 2$:
\begin{equation}
B\left(i\right) = \frac{i}{\ln\left(i-1/2\right)-\ln\left(1/2\right)}\,.
\end{equation}
From the last equations we see that the correction factor $\beta$, introduced above reads, e.g. for the case
of $\alpha \neq 2$, $\beta =2^{2-\alpha}$.  In the case of exponential distribution we get analogously:
\begin{eqnarray}
  kx_k(i) &=& B\left(i\right)\int_{k-1/2}^{k+1/2} k_1\exp\left(-k_1\right)dk_1
\\ &=& B\left(i\right)\left[ (k+1/2)e^{1/2-k}-(k+3/2)e^{-1/2-k} \right] \nonumber\,.
\end{eqnarray}
Here $B(i) =i/I_0$, with
\begin{eqnarray}
I_0(i) &=& \int_{1/2}^{i-1/2}k e^{-k} dk = (3/2) e^{-1/2} -(i+1/2)e^{1/2-i} \,. \nonumber
\end{eqnarray}

\begin{figure}
\centerline{\includegraphics[width=0.5\textwidth,angle=0]{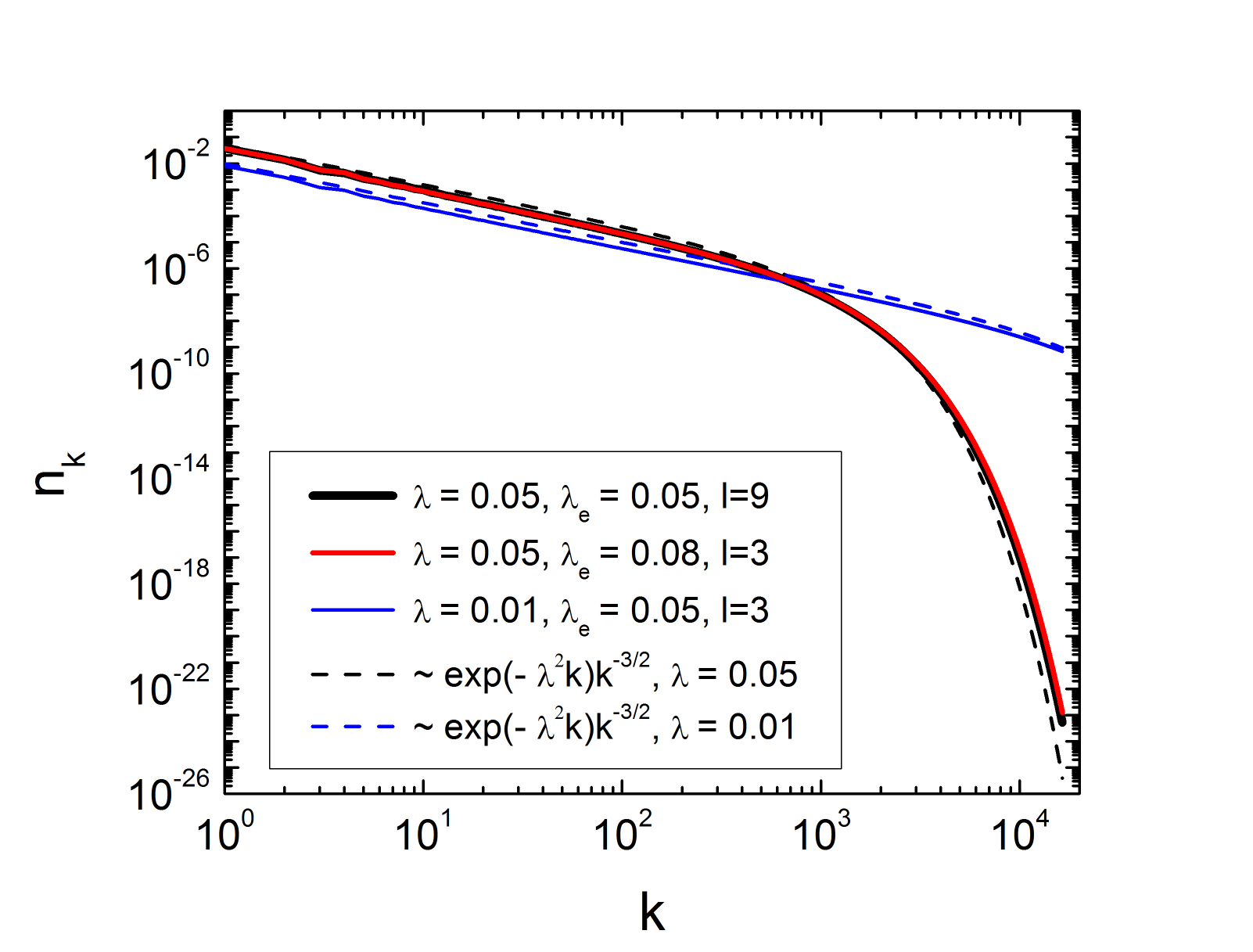}}
\caption{ {\bf Steady-state particle size distribution for completely disruptive collisions in the presence of collisions with an erosion. } The solid lines depict the size distribution for the systems with the following parameters: $\lambda=0.05$, $\lambda_e=0.05$, $l=9$ (black), $\lambda=0.05$, $\lambda_e=0.08$, $l=3$ (red), and $\lambda=0.01$, $\lambda_e=0.05$, $l=3$ (blue). The dashed lines indicate the respective size distribution for purely disruptive collisions for $\lambda=0.05$ (black) and $\lambda =0.01$ (blue).  All curves correspond to the case of constant kinetic coefficients. As it may be seen from the figure, the presence of the erosive collisions has practically negligible impact on the steady-state size distribution. } \label{witheros}
\end{figure}

\rule{0cm}{0.1cm}

\end{article}